\RequirePackage{fix-cm}

\documentclass[twocolumn]{svjour3}

\smartqed

\usepackage{hyperref}
\usepackage{graphicx}
\usepackage{mathtools}
\usepackage{subfig}
\usepackage{multirow}
\usepackage[font=small]{caption}
\usepackage[numbers,sort&compress]{natbib}

\usepackage{placeins}
\usepackage{accents}
\usepackage{stfloats}

\usepackage{tikz}
\usetikzlibrary{positioning, calc, fit, shapes, arrows, patterns}
\usetikzlibrary{shapes.geometric}
\usepackage[percent]{overpic}

\makeatletter
\tikzset{
        hatch distance/.store in=\hatchdistance,
        hatch distance=5pt,
        hatch thickness/.store in=\hatchthickness,
        hatch thickness=5pt
        }
\pgfdeclarepatternformonly[\hatchdistance,\hatchthickness]{north east hatch}
    {\pgfqpoint{-1pt}{-1pt}}
    {\pgfqpoint{\hatchdistance}{\hatchdistance}}
    {\pgfpoint{\hatchdistance-1pt}{\hatchdistance-1pt}}%
    {
        \pgfsetcolor{\tikz@pattern@color}
        \pgfsetlinewidth{\hatchthickness}
        \pgfpathmoveto{\pgfqpoint{0pt}{0pt}}
        \pgfpathlineto{\pgfqpoint{\hatchdistance}{\hatchdistance}}
        \pgfusepath{stroke}
    }
\makeatother

\makeatletter
\def\NAT@def@citea{\def\@citea{\NAT@separator}}
\makeatother

\newcommand{\RR}{\mathcal{R}}
\newcommand{\F}{F}
\newcommand{\cc}{C}
\newcommand{\multu}{q}
\newcommand{\ADJu}{\lambda_u}
\newcommand{\ADJmultu}{\lambda_\multu}
\newcommand{\norm}[1]{\left\lVert #1 \right\rVert}
\renewcommand{\det}[1]{\left| #1 \right|}

\newcommand{\void}{\mathrm{v}}
\newcommand{\solid}{\mathrm{s}}

\newcommand{\diff}[2]{\frac{\partial #1}{\partial #2}}

\newcommand{\ramp}[1]{\mathcal{E}\big(#1\big)} 
\newcommand{\rampder}[1]{\mathcal{E}'\big(#1\big)}

\newcommand{\designvar}{\chi}
\newcommand{\designvarold}{\chi_\mathrm{old}}
\newcommand{\lock}[1]{\mathcal{I}_{\designvar_\void}\!\big(#1\big)}
\newcommand{\lockder}[1]{\mathcal{I}'_{\designvar_\void}\!\big(#1\big)}

\newcommand{\kint}{k_i} 
\newcommand{\kreg}{k_r} 
\newcommand{\kregdimless}{\bar{k}_r}
\newcommand{\kvol}{k_\rho} 

\newcommand{\PsiNH}{\overline{\Psi}}
\newcommand{\scriptD}{\scriptscriptstyle D}
\newcommand{\scriptj}{\scriptscriptstyle\left<j\right>}
\newcommand{\scriptnum}[1]{\scriptscriptstyle\left<#1\right>}

\DeclareMathOperator{\dev}{dev}

\usepackage{stackengine}
\usepackage[export]{adjustbox}
\usepackage{enumitem}
\usepackage[bb=boondox,bbscaled=0.98]{mathalfa}
\DeclareMathOperator{\Hessian}{\mathbbb{H}\!}
\newcommand{\tensorm}{%
\setstackgap{S}{0.45ex}%
\mathrel{\Shortstack{{.} {.} {.}}}
}

\journalname{}

\begin{document}

\title{Internal contact modeling for finite strain topology optimization} 

\author{Gore Lukas Bluhm \and
        Ole Sigmund \and
        Konstantinos Poulios
}
\institute{Gore Lukas Bluhm \at \email{glubluh@mek.dtu.dk} \and
Department of Mechanical Engineering\\ 
Technical University of Denmark\\
Nils Koppels All\'{e}, Building 404\\ 
2800 Kgs. Lyngby, Denmark\\
}

\date{Received: date / Accepted: date}

\maketitle

\begin{abstract}
The present work proposes an extension of the third medium contact method for solving structural topology optimization problems that involve and exploit self-contact.
A new regularization of the void region, which acts as the contact medium, makes the method suitable for cases with very large deformations.
The proposed contact method is implemented in a second order topology optimization framework, which employs a coupled simultaneous solution of the mechanical, design update, and adjoint problems.
All three problems are derived and presented in weak form, and discretized with finite elements of suitable order.
The capabilities and accuracy of the developed method are demonstrated in a topology optimization problem for achieving a desired non-linear force-displacement path.
\end{abstract}

\section{Introduction}
In structural topology optimization under large deformations, it is not uncommon that beam-like members eventually come into contact.
Modeling of self-contact is therefore essential in order to avoid non-physical self-penetration in the mechanical analysis and thereby extend the accuracy and validity range of such optimization solutions.
Even more than that, a self-contact aware optimization can lead to radically different designs by exploiting the possibility of unilateral contact between parts of the design in e.g. non-linear mechanism design.

Applying standard methods for modeling of self-contact to situations like this, where the actual contact interfaces are a priori unknown, is in principle possible but also very tedious \cite{kumar2019a}.
Applying the third medium contact method \cite{wriggers2013a,2015Bo-Za-Ko-Ra}, instead, is a very natural and simple choice because density based topology optimization already involves a void region which can play the role of the contact medium.
Still, there are important technical aspects that require a careful treatment in order to make the combination numerically stable.

The third medium contact method was introduced in \cite{wriggers2013a} as a purely continuum mechanics based alternative to contact domain methods \cite{2009Ol-Ha-Ca-We-He}.
Its main advantage against standard contact methods, such as augmented Lagrangian and penalty ones \cite{1992Si-La,2010Gi-Po-Ge-Wa}, is that it avoids an explicit treatment of the contact kinematics and the implementation of inequality constraints.
However, it comes with the disadvantages of i) parasitic remote forces transferred between the bodies before the actual contact, and ii) not permitting large amounts of sliding between the contacting bodies.
Due to these limitations, third medium contact has hitherto not found significant applications as a replacement to the established contact methods based on Lagrange multipliers.

When it comes to topology optimization though, the third medium contact approach is an excellent fit, because density based topology optimization already employs a void phase in the a priori unknown regions where contact may actually occur.
By appropriate constitutive modeling, the void phase can easily be used as the contact medium.
At the same time, the inherent limitation of the method in capturing a sharp transition from zero to finite contact stresses, turns into an advantage in topology optimization.
The parasitic interaction of non-contacting parts of the structure due to small but finite stiffness of the third medium, makes the problem differentiable and hence allows the optimization to drive a design towards a self-contact state, if this is beneficial for the objective being optimized.

The idea of employing third-medium contact methods in topology optimization for modeling internal and external contact has so far been unexplored.
Nevertheless, traditional modeling of, mainly external, contact in topology optimization has been pursued in a series of papers.
Apart from earlier works in the field, reviewed by Hilding et al. \cite{1999Hi-Kl-Pe}, more recent works within small deformations topology optimization under external contact with rigid obstacles, include compliance or volume minimization in a frictionless \cite{2010St,2010St-Kl} or frictional \cite{2013St} setting, as well as compliance or contact pressure minimization in a frictional setting \cite{2020Kr-Po-Aa}.
Regarding large deformations topology optimization, Luo et al. \cite{2016Lu-Li-Ka} performed compliance minimization under frictionless contact with circular rigid obstacles, modeled by means of nonlinear springs.
Fernandez et al. \cite{2020Fe-Pu-So-To} used conventional large deformation contact methods to model external contact between the optimized design and a prescribed external deformable solid.
Known efforts for accounting for internal contact in compliance minimization problems are limited to a priori fixed internal contact interfaces, either without \cite{2007De} or with friction \cite{2020Ni-Zh-Ga}.
Otherwise, the level set method has also been successfully used for internal contact, though limited to zero initial gap and small sliding \cite{2015La-Ma}.

Apart from its simplicity, the major advantage of third medium contact in the context of density based topology optimization is that it completely eliminates the need for prescribing potential internal or external contact interfaces.
The main challenge lies in defining a constitutive law for the contact medium that is numerically stable even at extremely severe deformations and large sliding.
Apart from the currently employed constitutive laws for the void phase in large deformation topology optimization \cite{yoon2005a,2014Wa-La-Si-Je}, specialized third medium constitutive laws from the contact mechanics literature \cite{wriggers2013a,2015Bo-Za-Ko-Ra} provide a foundation for further developments.
The original model from \cite{wriggers2013a} with an explicitly added anisotropic stiffness term, has already been simplified in \cite{2015Bo-Za-Ko-Ra}, based on the fact that a highly distorted hyperelastic continuum actually exhibits the desired anisotropy naturally.
In the present work, a similar neo-Hookean material is used for the contact medium as the Hencky material used in \cite{2015Bo-Za-Ko-Ra}.
In addition, in order to allow for contact situations with considerable sliding and very distorted elements, the present work proposes a new void regularization based on penalization of higher order strains in void elements.

The basic idea can be basically implemented in any density based topology optimization algorithm accounting for large strains \cite{2014Wa-La-Si-Je,2015Wa-Ri}, minimizing a given objective function under a set of given constraints.
However, the present work introduces and uses a new second order topology optimization framework, fully defined in the continuous setting prior to discretization.
The approach is similar to the continuous adjoint sensitivity analysis by Cho and Jung \cite{cho2003a}, and is presented in a rather generalized form before being particularized to a specific hyperelastic material law and the objective function considered in the numerical example.
The included numerical results were obtained with the displacement field approximated with quadratic quadrilateral elements and the density field approximated with linear quadrilateral elements in a structured mesh but the actual method and its implementation is neither limited to structured meshes nor bound to a specific element type.

The subsequent section~2 presents some important adaptations of the third medium contact method that extend its applicability to larger amounts of sliding.
The topology optimization framework for the present work is introduced in section~3 and section~4 presents an optimization example that exploits the occurrence of self contact for achieving a desired nonlinear force-displacement response.
Important conclusions are summarized in section~5, while Table~\ref{tab:notation} summarizes notation conventions used throughout the present work.

\begin{table}[h!]
  \caption{Notation conventions}
  \label{tab:notation}
  \centering
  \begin{tabular}{l p{5cm}}\hline
  $A_{,x}(x;\delta x)$ & Gateaux differential of tensor $A$ with respect to tensor $x$ in direction $\delta x$ \\
  $A \cdot B = A_i B_i$ & Scalar product \\
  $A:B = A_{ij}B_{ij}$ & Double contraction \\
  $A\tensorm B = A_{ijk}B_{ijk}$ & Triple contraction \\
  $\det{A}$ & Determinant of matrix $A$ \\
  $\norm{A}=\sqrt{A:A}$ & Frobenius norm of matrix $A$ \\[5pt]
  $\nabla A = \dfrac{\partial A_i}{\partial X_j}$ & Spatial gradient of a vector field $A$ \\[10pt]
  $\Hessian A = \dfrac{\partial^2 A_i}{\partial X_j\partial X_k}$ & Spatial Hessian of a vector field $A$\\
  $\dev(A)$ & 3D deviator operator\\
  $\left< x \right>=\max(0,x)$ & Positive part function
\\ \hline
  \end{tabular}
\end{table}

\vspace{-2pt}
\section{Mechanical model}
The modeling domain $\Omega$ consists of a solid structure subdomain $\Omega_\solid$ embedded in a void phase subdomain $\Omega_\void$.
A vector field $u$ defined in the undeformed domain $\Omega$ expresses displacements from the initial to the current deformed configuration.
Let also $\multu$ denote a Lagrange multiplier vector which can either be a single vector or a vector field over some part $\Gamma_{\scriptD}$ of the domain boundary $\partial\Omega$.
In the former case, it is used to enforce a Dirichlet boundary condition in an average sense, while in the latter case, the condition is enforced at every point of $\Gamma_{\scriptD}$.
Based on these definitions, the mechanical equilibrium of the system can be expressed in weak form as
\begin{equation}
\RR(u,\multu;\,\delta u,\delta\multu) =0~~~~\forall \delta u,\delta\multu
\label{eq:equilibrium}
\end{equation}
with
\begin{equation}
\begin{split}
\RR(u,\multu;\,\delta u,\delta\multu)&=
\int_\Omega \Psi_{,u}(u;\,\delta u) ~d\Omega \\
&+ \int_{\Gamma_{\scriptD}} \Big\{\multu \cdot \delta u + (u-u_{\scriptD})\cdot\delta\multu\Big\} ~d\Gamma
\label{eq:equilibriumresidual}
\end{split}
\end{equation}
where $\Psi_{,u}(u;\,\delta u)$  is the virtual variation of the strain energy density function $\Psi$ due to an infinitesimal displacements variation $\delta u$.
This is the usual virtual work expression that can be rewritten in many equivalent forms of work conjugate pairs of strain variations and stresses.
Depending on whether the Dirichlet condition on $\Gamma_{\scriptD}$ is applied in an average sense or pointwise, $u_{\scriptD}$ is either a single vector or a  vector field expressing the prescribed displacement or displacement field.

\subsection{Hyperelasticity}
Regarding the strain energy function $\Psi$, there are many possible choices for hyperelastic material laws.
Most of these are formulated so that $\Psi$ tends to infinity when the material is compressed towards zero volume.
This property, apart from representing the actual physical response of any real material, is also essential for the third medium contact model employed here.

A rather common isotropic neo-Hookean material law according to Simo et al. \cite{simo1985a} is adopted in the present work both for the solid and the void domain, defined through the strain energy density function
\begin{equation}
\PsiNH(u) = 
\dfrac{K}{2} \left( \ln{\det{\F}} \right)^2  + \dfrac{G}{2} \left( \det{\F}^{-2/3} \norm{\F}^2 - 3 \right)
\label{eq:strain_energy_density}
\end{equation}
with the deformation gradient $\F=I+\nabla u$.
The initial bulk modulus $K$ and shear modulus $G$ depend on the material domain, being respectively equal to $K_\solid$ and $G_\solid$ in $\Omega_\solid$, and $K_\void$ and $G_\void$ in $\Omega_\void$.
These elasticity constants in the void domain $\Omega_\void$ are very small but finite.

Even if the void stiffness is initially negligible compared to the solid stiffness, the presence of the $\ln{\det{\F}}$ term in Eq.~\eqref{eq:strain_energy_density} will eventually lead to an infinite stiffness when the void is compressed to zero volume.
This property is crucial for the present application, because it results in a compressed void which is stiffer than the solid phase and can therefore transfer forces between solid members of the structure.
In general, every material law with this characteristic can serve as a third medium for modeling contact, in contrast to material laws, like the otherwise frequently used Saint Venant-Kirchhoff model, that result in vanishing stresses under ultimate compression.

The virtual work expression appearing in Eq.~\eqref{eq:equilibriumresidual}, results from Gateaux differentiation of $\Psi$ with respect to the displacements field $u$.
For the hyperelastic strain energy density from Eq.~\eqref{eq:strain_energy_density}, this operation results in
\begin{equation}
\PsiNH_{,u}(u;\,\delta u) = P\big(\nabla u\big) : \nabla \delta u
\label{eq:virt_work}
\end{equation}
with the 1\textsuperscript{st} Piola-Kirchhoff stress tensor
\begin{equation}
P\big(\nabla u\big) 
= K \ln\det{\F} \F^{-T}
+ G \det{\F}^{-2/3} \dev(\F\F^T)\F^{-T}
\label{eq:PK1}
\end{equation}

All equations have been presented in their three dimensional form, however a reduction to plane strain is trivial by defining the three dimensional deformation gradient as
\begin{equation}
\F = I+
\begin{bmatrix}\multicolumn{2}{c}{\nabla u~} & \begin{array}{c}0\\0\end{array}\\
                0 & 0 & 0
\end{bmatrix}.
\end{equation}
At the same time, for plane strain problems, $P$ in Eq.~\eqref{eq:virt_work} needs to be reduced to its in-plane components.

\subsection{Void regularization}
In general, the constitutive behavior of the void is not essential for the mechanical system as long as stresses in the void are small enough to not affect the deformation of the solid significantly.
Nevertheless, a rigorous modeling of the void up to large strains is essential for avoiding numerical instabilities of the overall system.
This is especially true when the void is used as the third medium for contact modeling, where many elements will collapse to almost zero volume.

\begin{figure}
    \centering
    \resizebox{0.45\textwidth}{!}{\begin{tikzpicture}

\draw [fill=gray!90, draw=black] (0,0) rectangle (5,5);
\draw [fill=gray!30, draw=black] (0.5,0.5) rectangle (5,4.5);
\fill [pattern=north east hatch, hatch distance=1.5mm, hatch thickness=.5pt] (-0.35,-0.35) rectangle (0,5.35);
\draw [-] (0,-0.35) -- (0,5.35);

\node (L) at (2.5,-0.2) {\normalsize $L$};
\draw [|<-] (0,-0.2) -- (L);
\draw [->|] (L) -- (5,-0.2);

\node (H) at (5.7, 2.5) {\normalsize $H$};
\draw [|<-] (5.7, 0) -- (H);
\draw [->|] (H) -- (5.7,5);

\node (t) at (5.4, 0.25) {\normalsize $t$};
\draw [|<->|] (5.2, 0) -- (5.2,0.5);

\node (omega_s) at (3.75, 0.25) {\normalsize $\Omega_s$};
\node (omega_v) at (2.75, 2.5) {\normalsize $\Omega_\mathrm{v}$};


\node (gamma_u) at (4, 5.3) {\normalsize $\Gamma_{\scriptD}$};
\draw (gamma_u.east) to[out=0, in=90] (4.85,5);
\draw [line width=0.5mm] (4.75,5) -- (5,5);

\node (gamma_uh) at (1, 5.3) {\normalsize $\Gamma_{\scriptD h}$};
\draw (gamma_uh.west) to[out=180, in=30] (0,5);
\draw [line width=0.5mm] (0,0) -- (0,5);

\draw [<->] (-1.25,0.35) -- (-1.25,-0.15) -- (-0.75,-0.15);
\node [anchor=west] at (-0.82,-0.15) {\scriptsize $x$};
\node [anchor=south] at (-1.25,0.27) {\scriptsize $y$};

\end{tikzpicture}}
    \caption{C-shape structure with $t = 0.1L$ and a length of $\Gamma_{\scriptD}$ equal to $0.05L$.}
    \label{fig:c_shape}
\end{figure}
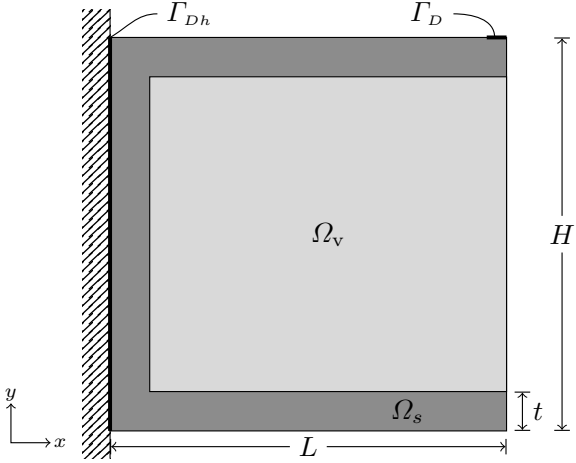

Figure~\ref{fig:c_shape} shows a classical benchmark example for dealing with large deformations in the void region \cite{yoon2005a,2014Wa-La-Si-Je}.
In the present work, the upper beam is loaded by prescribing the average vertical displacement within the small region $\Gamma_{\scriptD}$ through an incremented vertical component $u_{\scriptD y}$ in the vector $u_{\scriptD}$ in Eq.~\eqref{eq:equilibriumresidual}.
The horizontal displacement of the $\Gamma_{\scriptD}$ region is left free by defining $\multu\!=\!\left\{0,q_y\right\}^T$ and $\delta \multu\!=\!\left\{0,\delta q_y\right\}^T$ in Eq.~\eqref{eq:equilibriumresidual}, reducing the formulation to a scalar Lagrange multiplier  $q_y$, which is just a single additional degree of freedom in the overall system.
The homogeneous Dirichlet condition on $\Gamma_{\scriptD h}$ can be imposed by restricting the solution space of $u$.

\begin{figure}
    \centering
\subfloat
{
   \begin{tikzpicture}[every node/.style={anchor=south west,inner sep=0pt},
        x=1mm, y=1mm]
        \node (fig1) at (0,0)
        {\includegraphics[width=42mm, trim={390 395 1100 715}, clip]{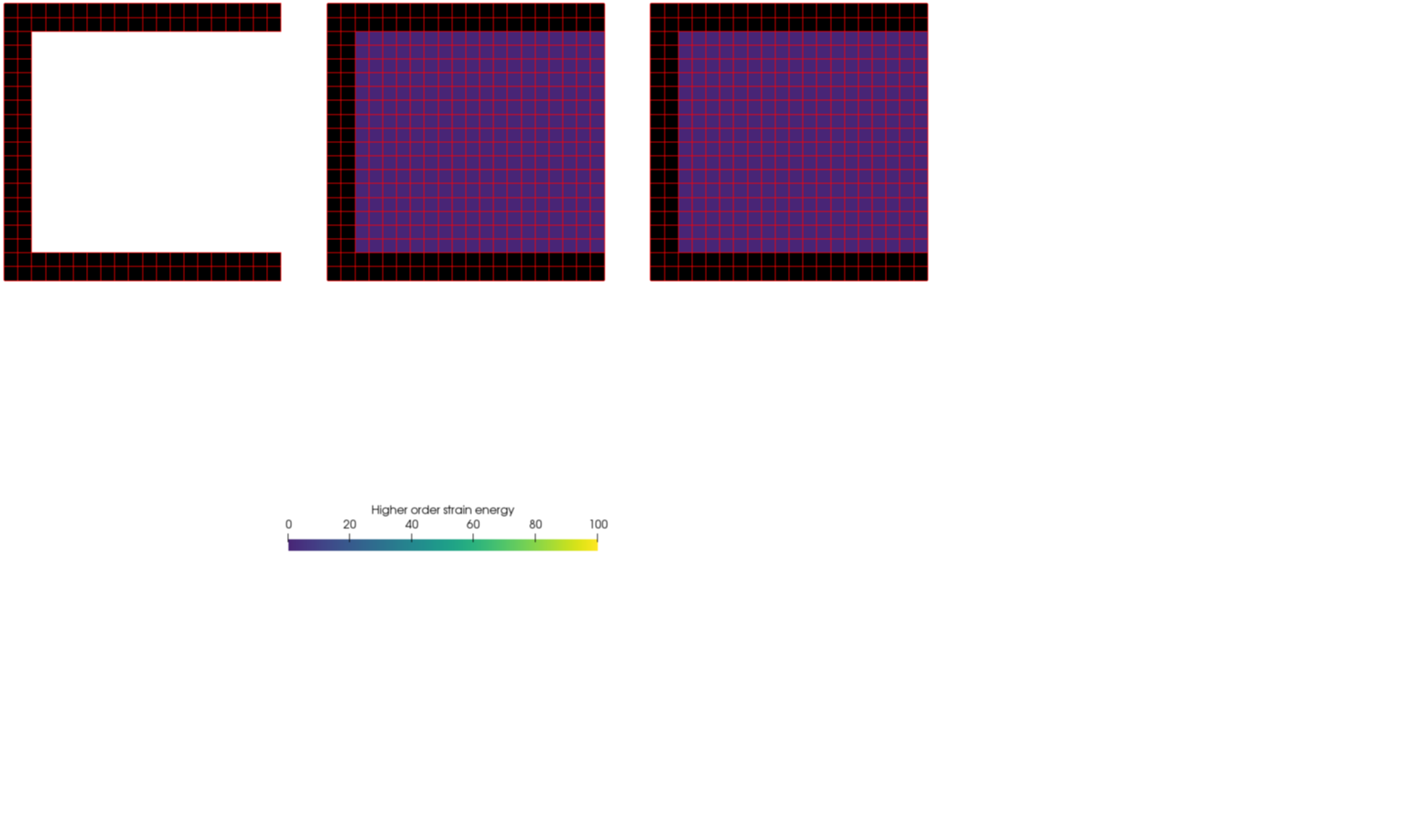}};
        \node at (3.2mm,6mm) {\scriptsize Higher order strain energy $\Tilde{\Phi}$};
\end{tikzpicture}
}
\hfil
\addtocounter{subfigure}{-1}
\hfil
\subfloat[$u_{\scriptD y} = -0.2H$ \label{fig:reg_struct_0.2}]
{\includegraphics[width=84mm, trim={0 770 650 0}, clip]{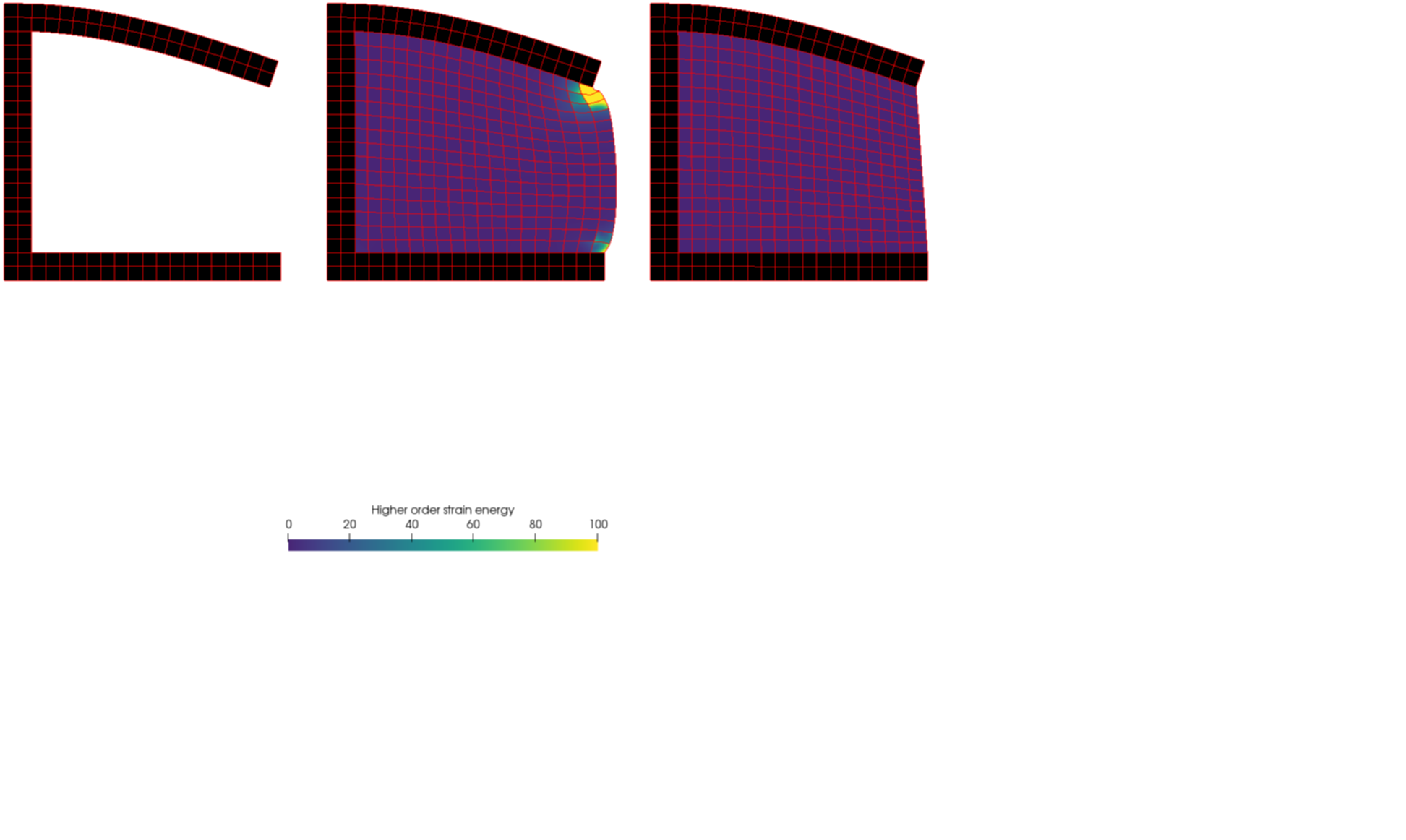}}
\hfil
\subfloat[$u_{\scriptD y} = -0.4H$ \label{fig:reg_struct_0.4}]
{\includegraphics[width=84mm, trim={0 770 650 0}, clip]{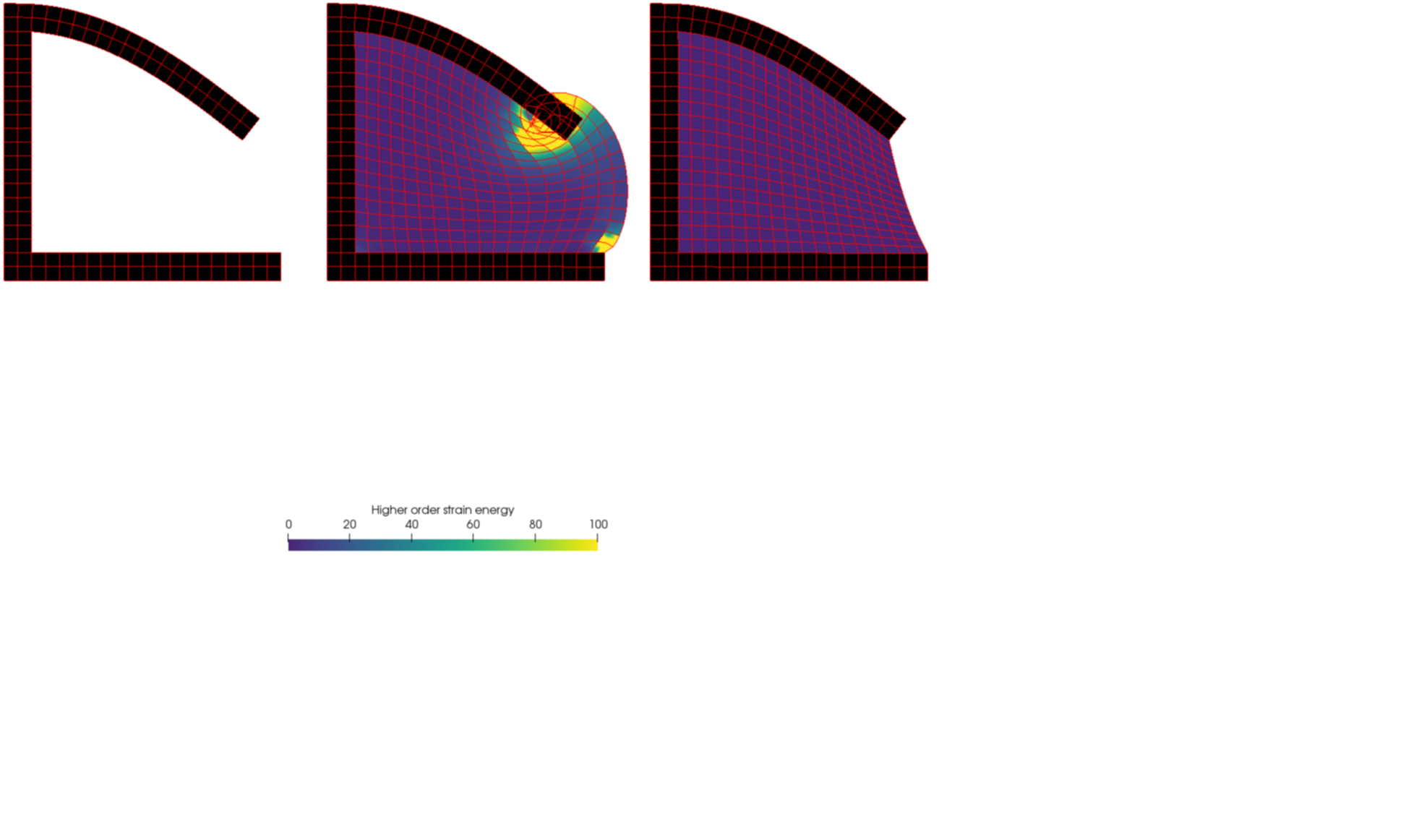}}
\hfil
\subfloat[$u_{\scriptD y} = -0.6H$ \label{fig:reg_struct_0.6}]
{\includegraphics[width=84mm, trim={0 770 650 0}, clip]{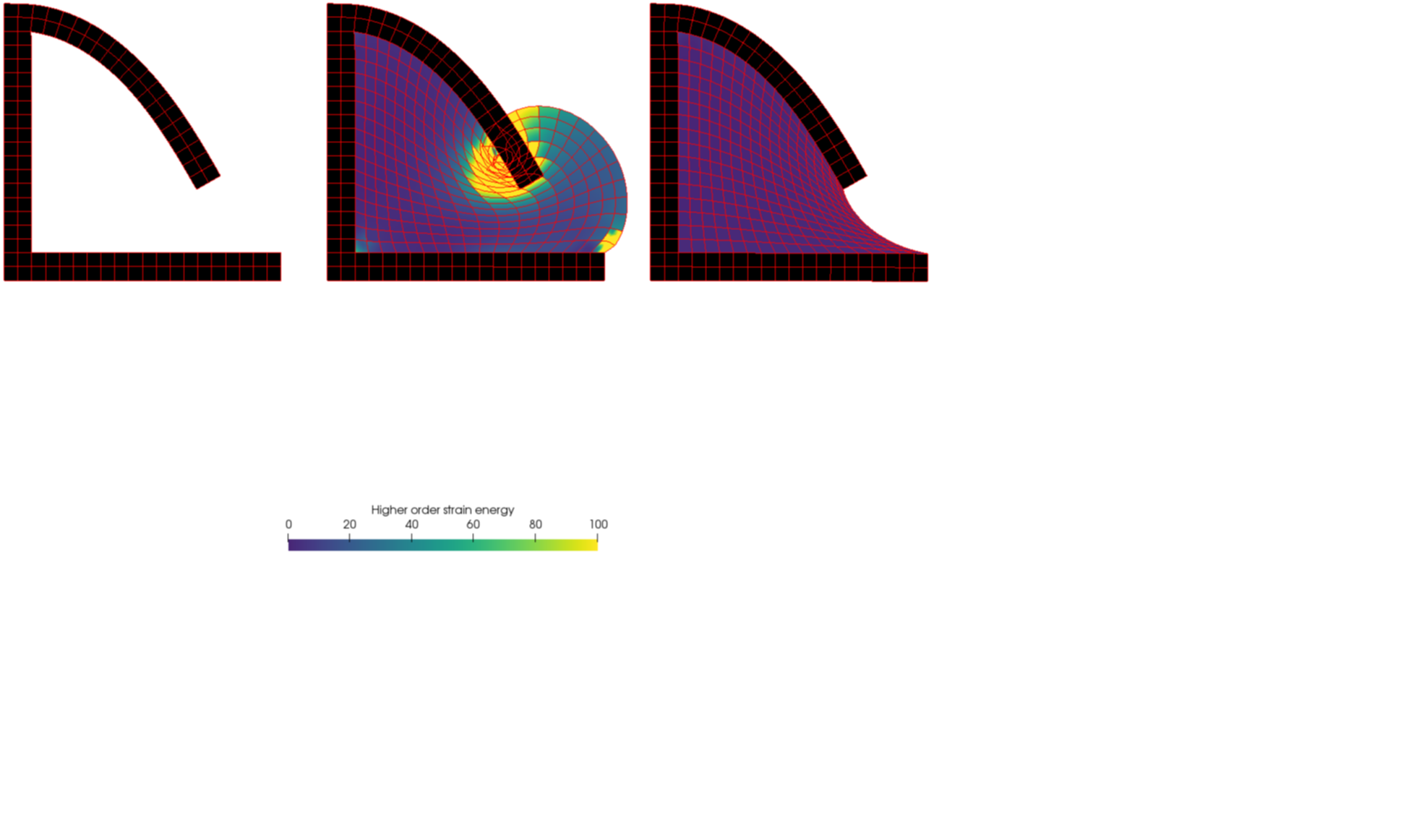}}
\caption{Deformed C-shape structure for $H=L=1$, with no void material (left), without void regularization (middle) and  with void regularization (right). Solid structure is shown in black, void is colored according to the higher order strain energy $\Tilde{\Phi}$ defined in Eq.~\eqref{eq:benging_energy}.}
\label{fig:reg_struct}
\end{figure}

The plots in the first column of Figure~\ref{fig:reg_struct} show simulation results for the reference solution with $K_\solid\!=\!1$ and $G_\solid\!=\!6/13$ in stress units, and the void elements completely removed.
The second column shows the corresponding results after including void elements with $K_\void/K_\solid\!=\!G_\void/G_\solid\!=\!10^{-12}$.
All results are obtained with a $20\!\times\!20$ discretization of the total domain $\Omega$ with \mbox{8-node} quadratic quadrilateral elements and numerical integration with nine Gauss points.
Although the simulation does not break down in the load range shown in Figure~\ref{fig:reg_struct}, the deformed void mesh in the second column has many severely deformed elements which are close to being inverted.
At higher load levels, inverted or severely distorted elements lead to numerical instabilities preventing the employed Newton algorithm from converging.

Different methods have been successfully applied for avoiding this kind of degenerate states of the deformed void \cite{yoon2005a,2014Wa-La-Si-Je}.
The present work proposes a new regularization which gives excellent control over the void mesh deformation and at the same time is compatible with the use of the void as a third medium for contact modeling.
The basic idea is to augment the material strain energy function $\PsiNH$ with an energy associated with higher order strains within each finite element according to an energy density proportional to the function
\begin{equation}
    \Tilde{\Phi} = \dfrac{1}{2} \Hessian{u} \tensorm \Hessian{u}
    \label{eq:benging_energy}
\end{equation}
where $\Hessian{u}$ is the Hessian of the displacement field, i.e. the second order strain tensor.
The colorplot shown in the second column of Figure~\ref{fig:reg_struct} illustrates large values of $\Tilde{\Phi}$ occurring in regions of severe mesh distortion.

In order to penalize these deformation modes, one could consider an augmented strain energy function \mbox{$\PsiNH+\kreg\,\mathcal{I}\,\Tilde{\Phi}$}, with the regularization scaling constant $\kreg$ and the void indicator function $\mathcal{I}$, which would lead to the virtual work expression
\begin{equation}
    \PsiNH_{,u}(u;\,\delta u)
    + \kreg\,\mathcal{I}\,\Hessian{u} \tensorm \Hessian{\delta u}.
    \label{eq:virt_work_aug_orig}
\end{equation}
This is a rather non-intrusive approach because it would only penalize bending and warping deformation modes in void elements, while it would not penalize homogeneous deformation states even at very severe homogeneous compression and shearing.

In order to reduce the impact of the regularization term further, an ad hoc scaling of the higher order term in Eq.~\eqref{eq:virt_work_aug_orig} is proposed, leading to the weak form
\begin{equation}
\Psi_{,u}(u;\,\delta u) = \PsiNH_{,u}(u;\,\delta u) + \kreg\,\mathcal{I} \,e^{-5\det{F}}\,\Hessian{u}\!\tensorm\!\Hessian{\delta u}
\label{eq:virt_work_aug}
 \end{equation}
to be used, with a slight exploitation in notation, in Eq.~\eqref{eq:equilibriumresidual}.
The added negative exponential dependency on the determinant of the deformation gradient is derived from numerical experiments.
It leads to a decaying intensity of the regularization for elements that are not severely compressed, or even less for elements that are stretched.

The penalization constant $\kreg$ has force units. In order to work with a dimensionless quantity with more general validity, the following scaling relationship is proposed
\begin{equation}
    \kreg = \kregdimless L^2 K_\solid.
\end{equation}
where $L$ is used as a characteristic length of the structure and $K_s$ is the elastic bulk modulus of the solid.
The dimensionless constant $\kregdimless$ has to be chosen just large enough to avoid extreme distortion of void elements.
Convergence studies with regard to the parameter $\kregdimless$ have led to the value $\kregdimless = 10^{-6}$ used in all numerical examples throughout the present work.
Note that the added Hessian term serves its purpose of regularizing the deformed element shape, even for finite element spaces for $u$ which are not $C^1$ continuous across element boundaries, as a compatible discretization of this term would require.

\subsubsection*{C-shape structure example}
The last column in Figure~\ref{fig:reg_struct} shows simulation results for the square C-shape structure with the void regularization according to Eq.~\eqref{eq:virt_work_aug} and otherwise the same void material parameters as the results of the second column.
Before adding the regularization, the void material was squeezed out at the open right boundary of the domain, resulting in a severely bent mesh, especially near the tip of the loaded beam.
With the proposed regularization, all void region elements remain completely regular, which is also reflected in the much lower bending and warping energy plotted in the void region.
When looking at the bottom beam at the highest load, a barely visible negative vertical deflection for the case with the regularized void, reveals a stronger parasitic interaction between the loaded and the unloaded beam.
This weak but still detectable effect is consequence of an added bending stiffness due to the regularization.

The last load step, shown in Figure~\ref{fig:reg_struct_0.6}, indicates how more prone to numerical instabilities the model without void regularization is, compared to the model with void regularization.
There exist other void regularization schemes in the literature which are as effective in this benchmark example.
For this reason, the additional potential of the proposed regularization is further illustrated in a modified version of this example, that involves contact between the two beams.

\begin{figure*}[t]
\centering
\subfloat[$u_{\scriptD y} = -0.2H$ \label{fig:double_c_contact_0.2}]
{\includegraphics[width=1.\textwidth, trim={0 890 260 0}, clip]
                 {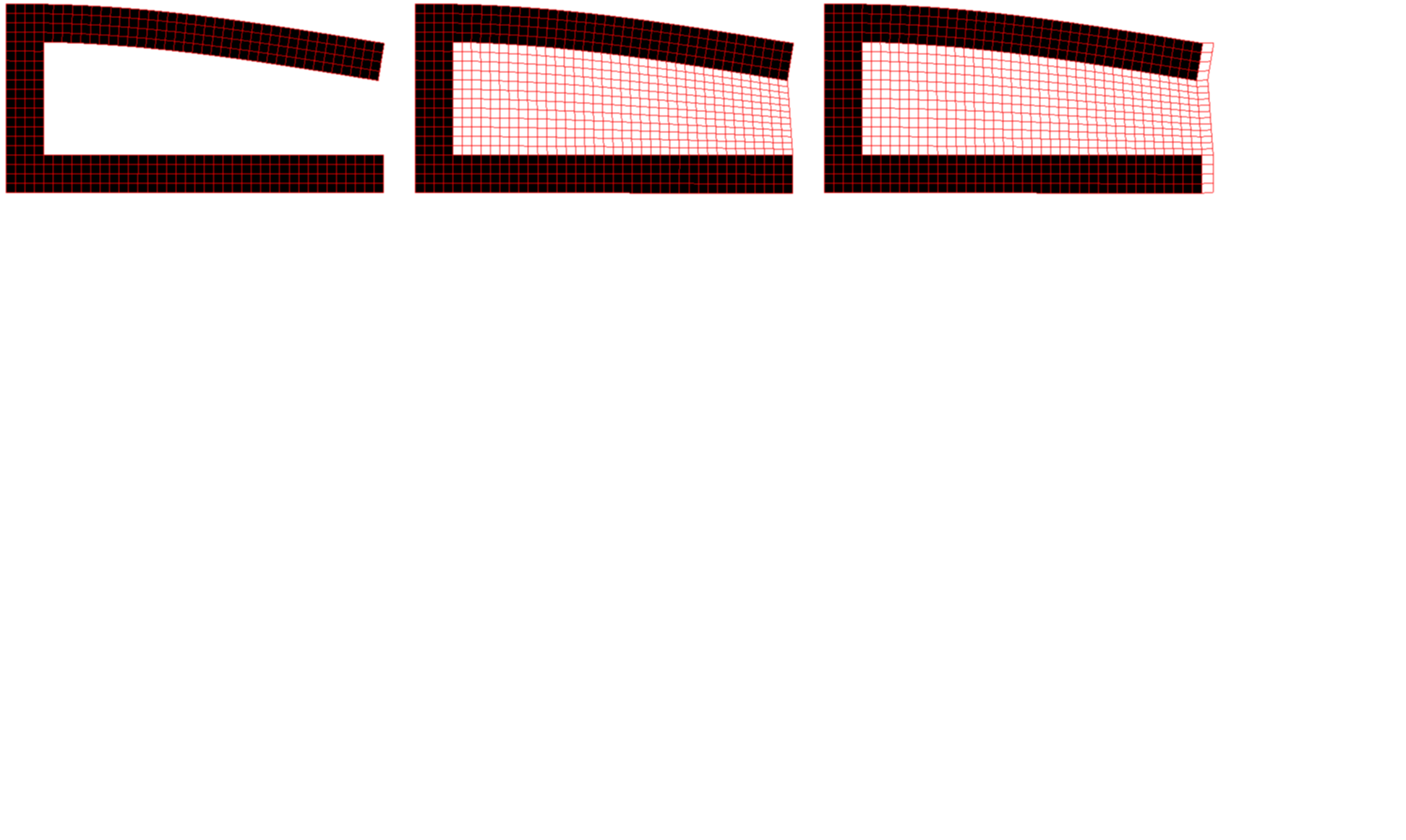}}\\[3pt]
\subfloat[$u_{\scriptD y} = -0.5H$]
{\includegraphics[width=1.\textwidth, trim={0 890 260 0}, clip]
                 {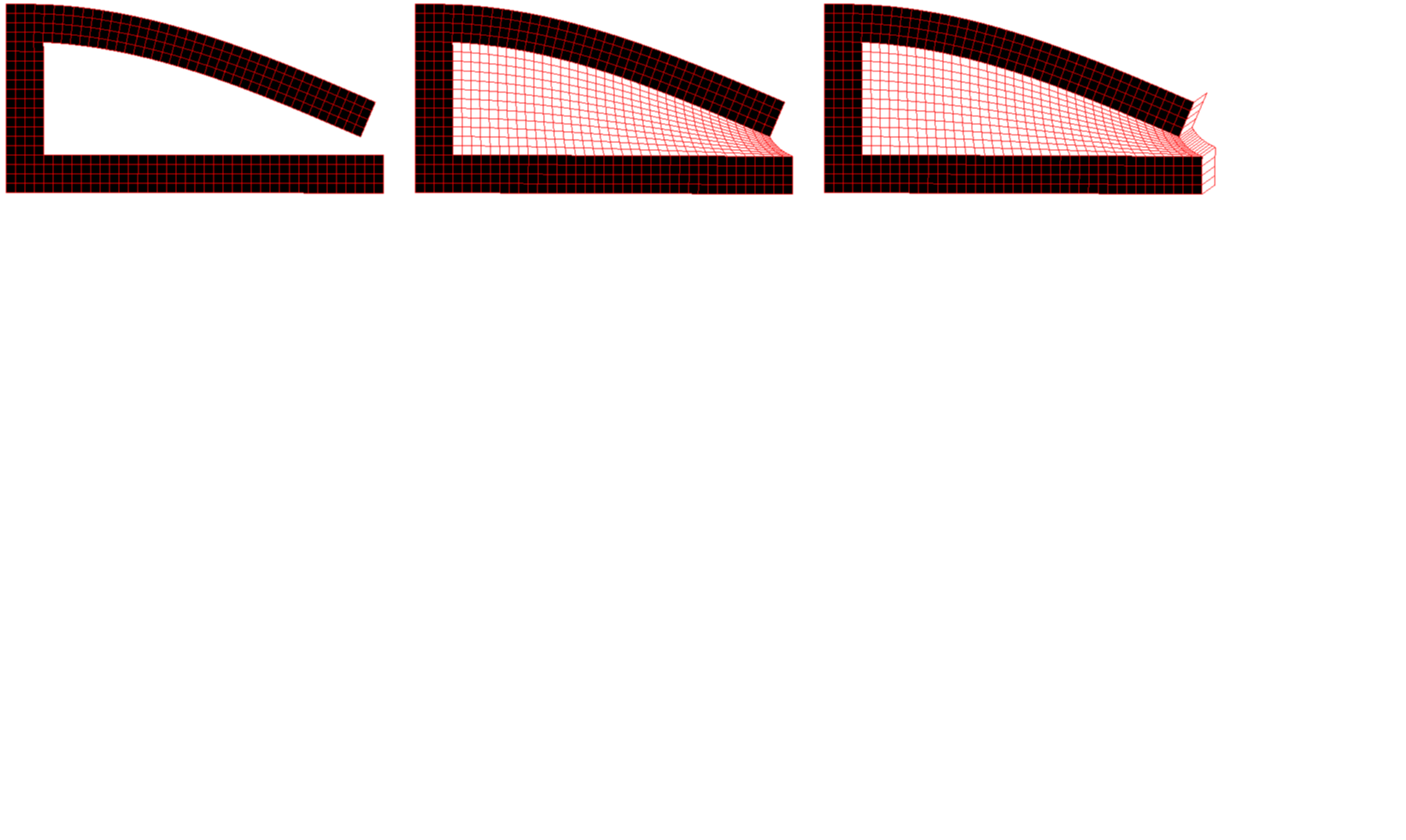}}\\[3pt]
\subfloat[$u_{\scriptD y} = -H$]
{\includegraphics[width=1.\textwidth, trim={0 760 260 0}, clip]
                 {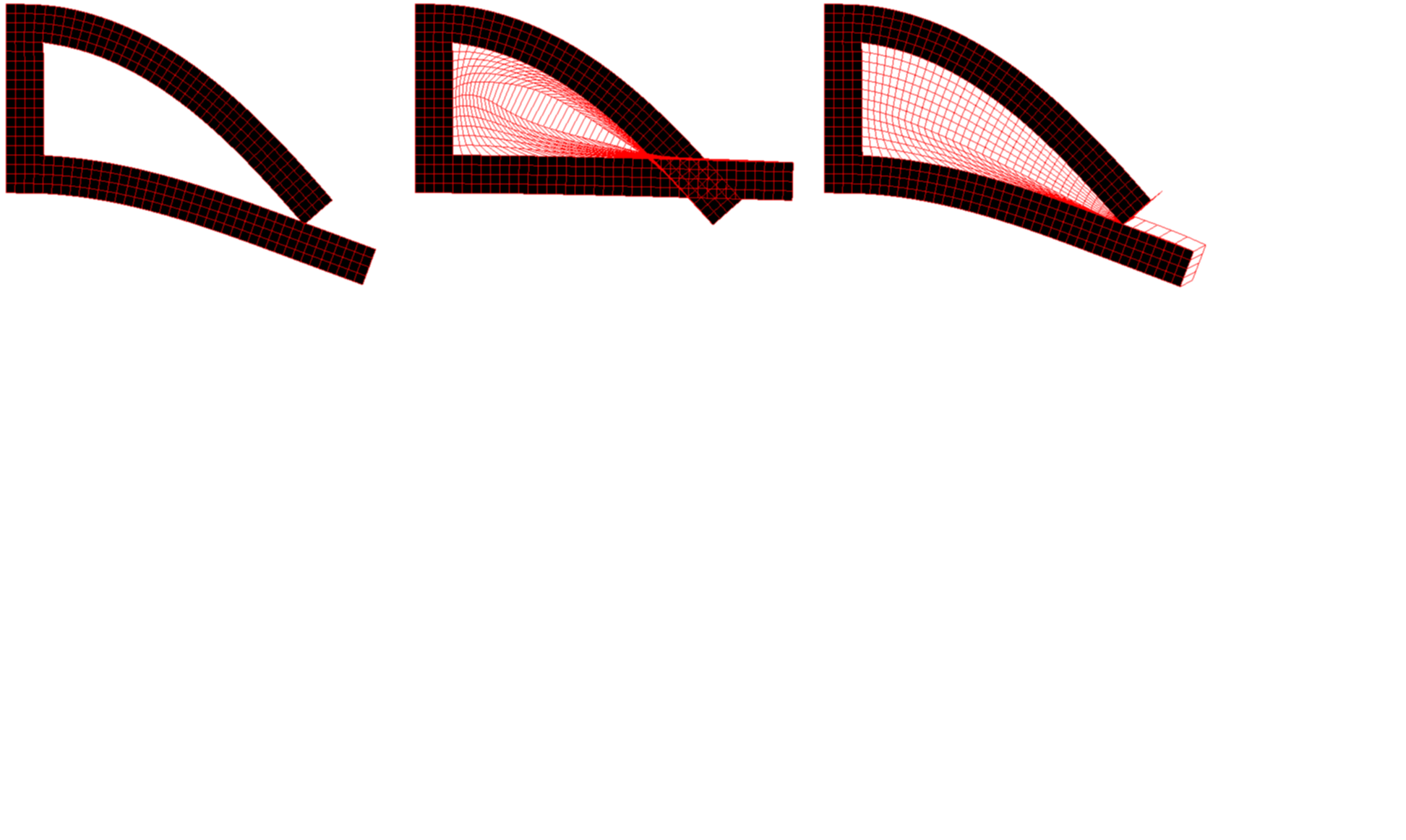}}
\caption{Deformations of the C-shape structure with $H = 0.5L$, with no void material solved using a conventional contact formulation based on Lagrange multipliers (left), with void regularization (middle) and  with void regularization and an additional layer of void material at the open right hand side boundary (right).}
\label{fig:double_c_contact}
\end{figure*}

\subsubsection*{Half height C-shape structure example}
In the next example, shown in Figure~\ref{fig:double_c_contact}, the height of the C-shape structure is reduced to half, i.e. $H\!=\!0.5L$, so that upon vertical loading of the upper beam, the two horizontal beams will eventually come into contact.
The effectiveness of the proposed void regularization in modeling self-contact is demonstrated by imposing a displacement of the upper beam up to $u_{\scriptD y}\!=\!-H$.
The structure is discretized with 40 times 20 elements along the length and height directions, respectively.

In fact, this is a rather challenging contact example, because contact will occur at the very corner of the upper beam against the top surface of the lower beam.
The first column of Figure~\ref{fig:double_c_contact} shows the deformed structure at different loading states, with no void material and conventional contact modeling with Lagrange multipliers \cite{2015Po-Re}, as a reference solution.
The second column shows results with the void between the two horizontal beams modeled with the void regularization according to Eq.~\eqref{eq:virt_work_aug}.
The void mesh remains very regular, as was the case in the previous example, but despite the absence of any inverted elements, the free edge at the right boundary of the domain folds into a configuration that leads to overlapping.
The fact that the expected contact point is at the boundary of the void domain prevents the void from acting as a contact medium between the two beams.

This is easy to overcome simply by extending the void domain with an additional column of elements at the right boundary of the domain, as shown in the last column of Figure~\ref{fig:double_c_contact}. In fact, when applying the third medium approach for contact, it is essential, that every potential contact region is entirely embedded in the contact medium.
With a void domain that sufficiently embeds the contact region in this case, the regularized void acts as a contact medium transferring forces and permitting penetration between the two beams even for the significant amount of sliding observed at the maximum imposed displacement $u_{\scriptD y}\!=\!-H$.
At a moderate displacement of the upper beam with $u_{\scriptD y}\!=\!-0.2H$, both simulations with the regularized void exhibit no visible deviations from the reference solution.
At the intermediate loading state with $u_{\scriptD y}\!=\!-0.5H$, a barely visible deflection of the lower beam indicates a small parasitic interaction between the two beams that have not yet come into contact.
At the post contact deformation state with $u_{\scriptD y}\!=\!-H$, the result seen in the right hand side column for the regularized void with the additional buffer layer of void is in good agreement with the reference solution.

\begin{figure}[h!]
    \centering
    \includegraphics[width=84mm]{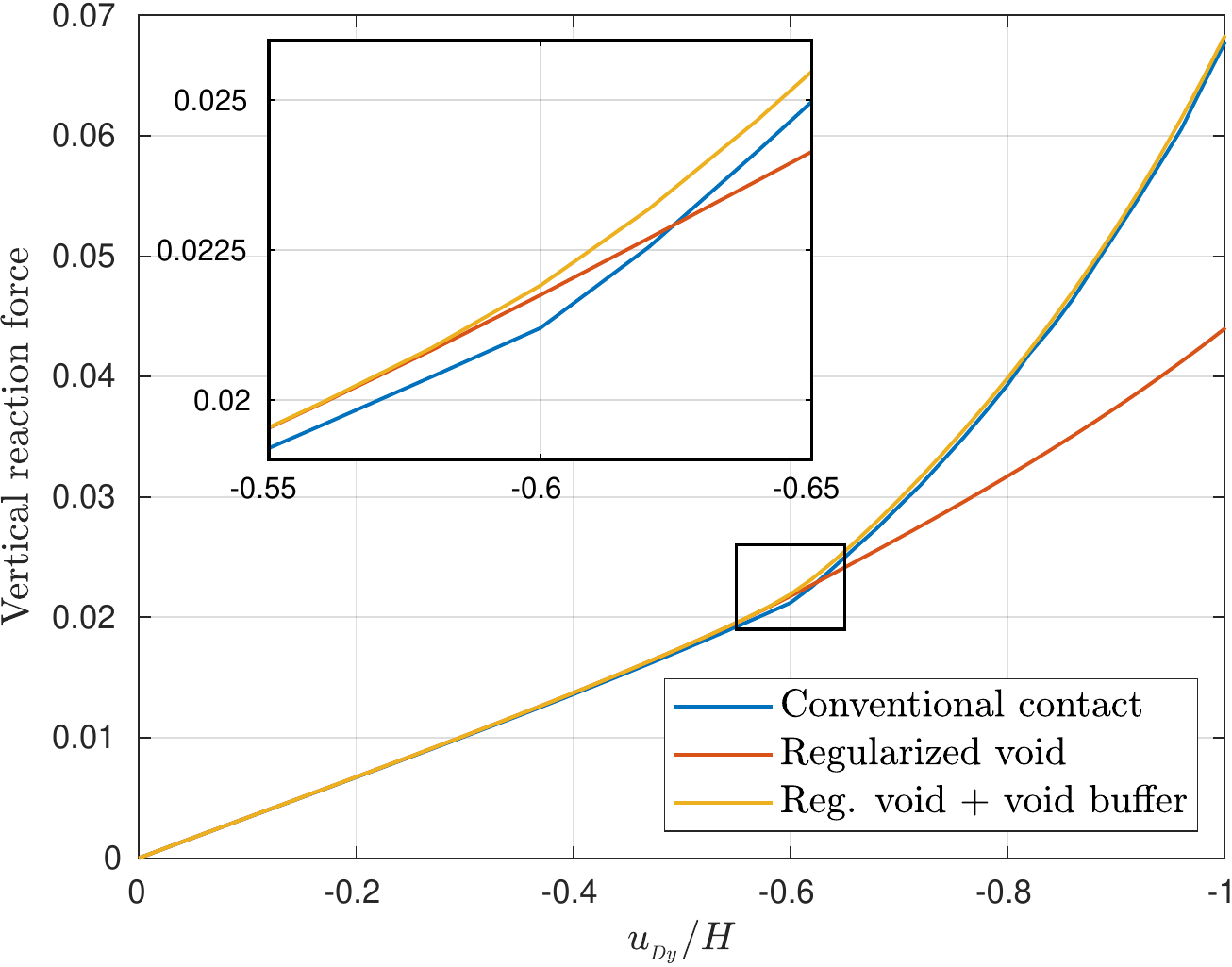}
    \caption{Load-displacement responses at $\Gamma_{\scriptD}$ for the three simulations of the C-shape structure from Figure~\ref{fig:double_c_contact}.}
    \label{fig:double_c_contact_comp}
\end{figure}

A more quantitative comparison is provided in Figure~\ref{fig:double_c_contact_comp} that illustrates the obtained load displacement curves for the three cases from Figure~\ref{fig:double_c_contact}.
The two curves for the simulations with void material coincide until the onset of contact and exhibit a just slightly larger reaction force compared to the reference case.
Beyond the occurrence of contact, the case with the appropriately extended void domain follows the reference solution with a very small deviation even at a significant amount of sliding.
The inset detail view of the transition point at the onset of contact helps estimating the loss of accuracy in predicting a sharp transition due to the inherent limitation of the third medium contact approach, compared to conventional contact modeling.

\subsection{Material interpolation}
\label{subsec:material_interpolation}
So far, only structures have been treated that involve solid and void domains clearly delimited by a discontinuous interface.
However, density based topology optimization relies on the concept of a continuous material density varying from void to solid in a differentiable manner.
A graded interface between the two regions is essential for establishing the sensitivity of an objective function with respect to changes of the design.
With this motivation, a material interpolation scheme is introduced below for representing a transition from solid to void regions, and the effectiveness of the proposed third medium contact is also evaluated in this context.

\begin{figure}[!ht]
\centering
\subfloat[Mapping: $\designvar \rightarrow \rho(\designvar)$\label{fig:projection}]
         {\includegraphics[scale=0.7]{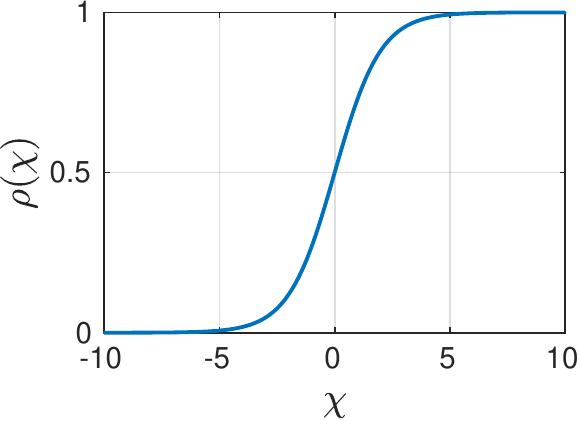}}~
\subfloat[RAMP ($p = 8$) \label{fig:ramp}]
         {\includegraphics[scale=0.7]{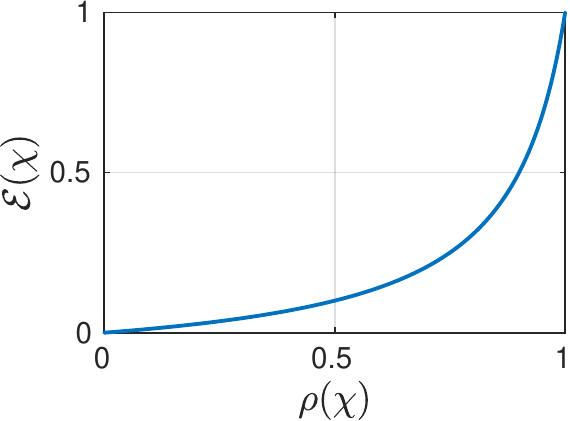}}\\
\subfloat[Void indicator at $\designvar_\void\!=\!-5$ \label{fig:void_indicator}]
         {\includegraphics[scale=0.7]{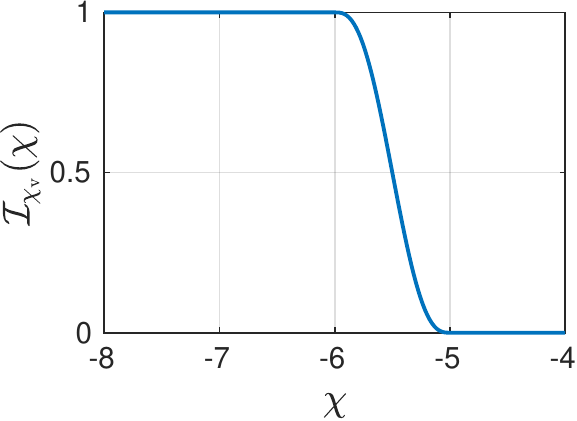}}~
\subfloat[All transition functions \label{fig:transition_functions}]
         {\includegraphics[scale=0.7, trim={0 0 0 0}, clip]{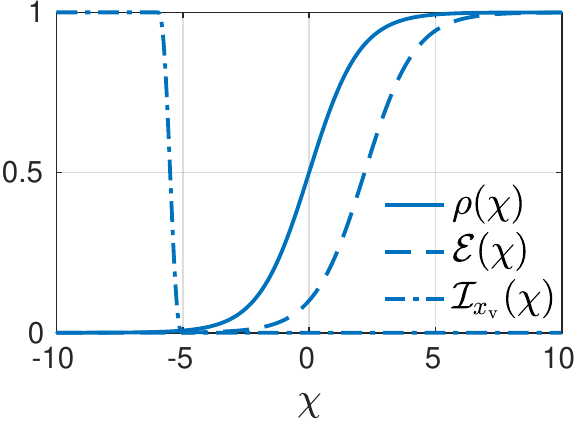}}
\caption{a) Mapping from the design variable to the physical density, Eq.~\eqref{eq:projection}. b) RAMP material stiffness interpolation function, Eq.~\eqref{eq:ramp}. c) Void indicator function, Eq.~\eqref{eq:void_indicator}, for activating void regularization. d) Comparison of material density, material stiffness and void regularization transition functions.}
\end{figure}

The design is described by a continuous design variable field $\designvar$ which represents a purely mathematical unconstrained quantity, i.e. $\designvar\in(-\infty,\infty)$.
To express material consumption, $\designvar$ is mapped to a physical material density $\rho$ through the function
\begin{equation}
    \rho(\designvar) = \dfrac{1}{1+\mathrm{e}^{-\designvar}},
    \label{eq:projection}
\end{equation}
which results in a density field within the asymptotic limits of 0 and 1, corresponding to a perfect void and a perfect solid material, respectively, as illustrated in Figure~\ref{fig:projection}.
According to this mapping, the conventional solid to void boundary at $\rho\!=\!0.5$ corresponds to \mbox{$\designvar\!=\!0$}, while for values of $\designvar$ in the order of $+10$ and $-10$, respectively, the deviation from a perfect solid and void material, has no practical significance.

Although the density field is continuous, the aim is nevertheless to achieve manufacturable and discrete designs where $\rho$ is either at its upper or lower bound, while intermediate densities are limited to the interface between solid and void domains.
Usually this is done by penalizing the stiffness of intermediate densities through a convex relationship between material usage $\rho$ and material performance, i.e. stiffness for structural topology optimization.
The present work adopts the RAMP material interpolation function by Stolpe and Svanberg \cite{stolpe2001a} due to its finite initial slope.
Expressed in terms of the mathematical design variable $\designvar$, this interpolation is represented by the function
\begin{equation}
    \ramp{\designvar} = \mathcal{E}_0 + \left(1-\mathcal{E}_0\right) ~ \frac{\rho(\designvar)}{1+p(1-\rho(\designvar))},
    \label{eq:ramp}
\end{equation}
where $p$ is the penalization parameter and $\mathcal{E}_0$ is a small positive value ensuring minimum stiffness even if the density is very close to zero.
The RAMP interpolation function from Eq.~\eqref{eq:ramp} is applied as scaling to the solid material elastic properties resulting in the initial bulk modulus $K\!=\!\ramp{\designvar} K_\solid$ and the initial shear modulus $G\!=\!\ramp{\designvar} G_\solid$.
A linear interpolation is obtained for $p\!=\!0$, while increasingly positive values for $p$ result in an increasing penalization of intermediate densities.
Figure~\ref{fig:ramp} illustrates the RAMP function for the penalization parameter $p\!=\!8$, used throughout the present work, while the compound function $\ramp{\designvar}$ for the same value of $p$ is shown in Figure~\ref{fig:transition_functions}.\\

In the context of a continuous material transition from void to solid, discussed in this subsection, the void indicator function $\mathcal{I}$, which acts as a switch for the void regularization introduced in Eq.~\eqref{eq:virt_work_aug}, needs also to be redefined in a differentiable manner.
In order to apply the regularization term only to regions where the design variable $\designvar$ is below a specified threshold $\designvar_\void$, the following smooth negative step function between $\designvar_\void-1$ and $\designvar_\void$, adopted from \cite{2003Eb-Mu-Pe-Pe-Wo-Ma-Ha}, can be used.
\begin{equation}
\begin{split}
    \lock{\designvar} =&~~6 \big<\designvar_\void-\designvar\big>_{[0,1]}^5\\
                       &-15 \big<\designvar_\void-\designvar\big>_{[0,1]}^4\\
                       &+10 \big<\designvar_\void-\designvar\big>_{[0,1]}^3
\end{split}
\label{eq:void_indicator}
\end{equation}
with
\begin{equation}
\left<\designvar\right>_{[0,1]}
    =\left\{\begin{array}{ll}0 & ~~\mathrm{if}~\designvar< 0\\
	                         1 & ~~\mathrm{if}~\designvar> 1\\
	                         \designvar & ~~\mathrm{if}~0 \leq \designvar \leq 1\\
             \end{array}\right.
\end{equation}
Figure~\ref{fig:void_indicator} illustrates this smooth indicator function for the threshold value $\designvar _\void\!=\!-5$, used throughout this work.
The function yields 0 for $\designvar\!\ge\!-5$ and 1 for $\designvar\!\le\!-6$, which means that the void regularization is completely absent in solid and intermediate density regions, as also seen in Figure~\ref{fig:transition_functions}.

After having introduced a dependence of the material elasticity parameters on the design variable $\designvar$ and after replacing $\mathcal{I}$ in Eq.~\eqref{eq:virt_work_aug} with the smooth void indicator function from Eq.~\eqref{eq:void_indicator}, the mechanical equilibrium residual according to Eq.~\eqref{eq:equilibriumresidual} becomes
\begin{equation}
\begin{split}
\RR(\designvar,u,&\multu;\,\delta u,\delta\multu)= \\
& \int_\Omega \Big\{\ramp{\designvar} P_\solid(\nabla u)\!:\!\nabla \delta u\\
&~~~~~~+ \kreg\,\lock{\designvar}\,e^{-5\det{F(\nabla u)}}\,\Hessian{u}\!\tensorm\!\Hessian{\delta u}\Big\} ~d\Omega \\
+& \int_{\Gamma_{\scriptD}} \Big\{\multu \cdot \delta u + (u-u_{\scriptD})\cdot\delta\multu\Big\} ~d\Gamma,
\end{split}
\label{eq:equilibriumresidual2}
\end{equation}
where $P_\solid$ is the 1\textsuperscript{st} Piola-Kirchhoff stress tensor from Eq.~\eqref{eq:PK1} for the completely solid material, i.e. $K\!=\!K_\solid$ and  $G\!=\!G_\solid$.

\subsubsection*{Half height C-shape structure with graded interface}

\begin{figure}[b!]
\subfloat[Undeformed\label{fig:cont_double_c_0}]{
\begin{tikzpicture}[every node/.style={anchor=south west,inner sep=0pt},x=1mm, y=1mm]
    \node (fig1) at (0,0)
    {\includegraphics[scale=0.227, trim={0 686 1220 0}, clip]
                     {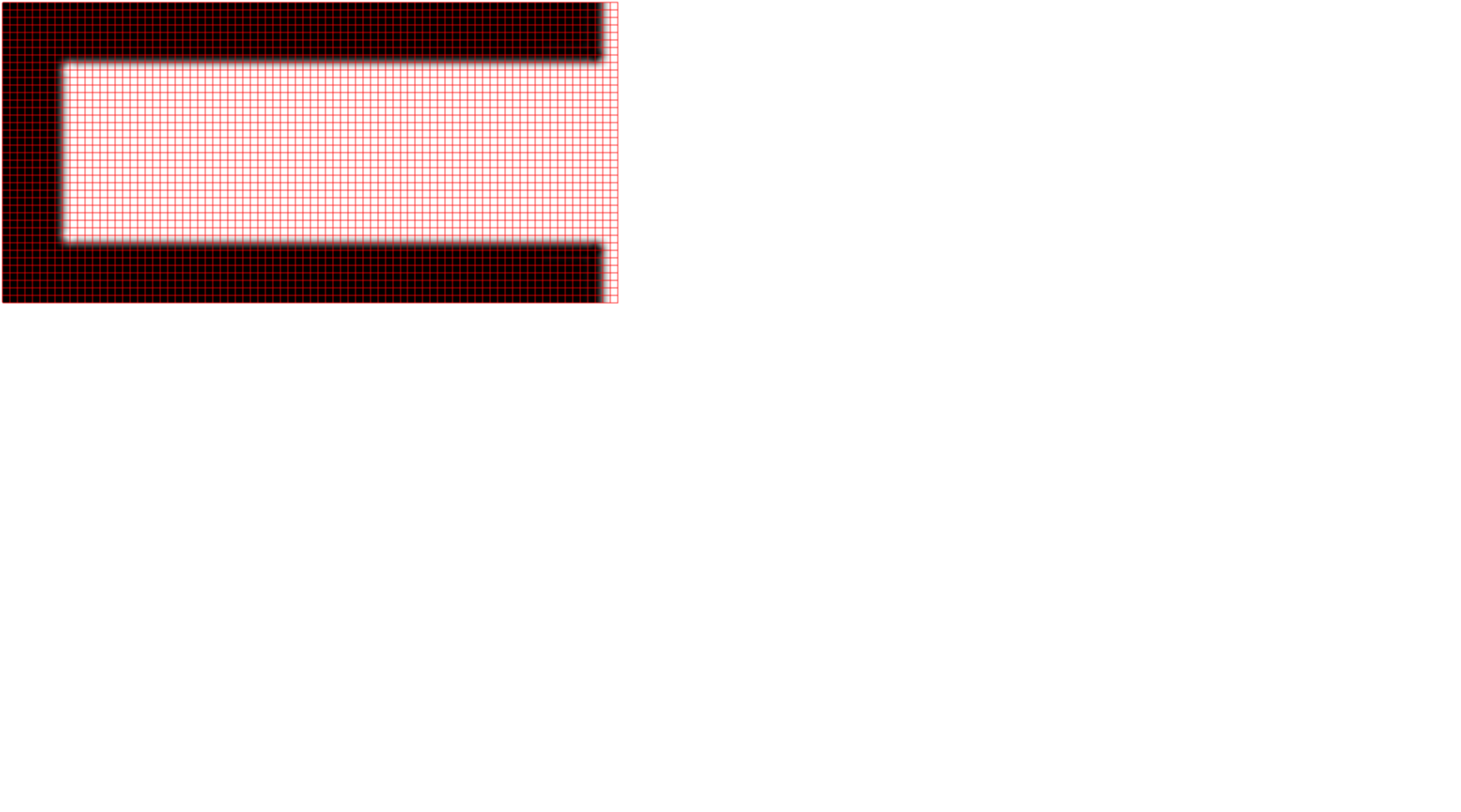}
    \raisebox{2.4mm}{
    \includegraphics[height=3.83cm, trim={0 18 38 0}, clip]
                    {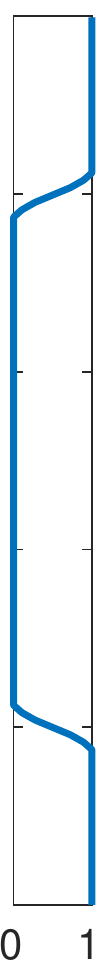}
    }
    };
    \node (omega) at (75.8mm, 0) {\footnotesize $\rho$};
    \draw[<->, line width=0.5mm, color={rgb:red,0;green,0.4470;blue,0.7410}] (65,5) -- (60,5) -- (60,40) -- (65,40);
\end{tikzpicture}
}\\[-5pt]
\subfloat[$u_{\scriptD y} = -0.5H$]
{\includegraphics[scale=0.231, trim={0 385 590 0}, clip]
                 {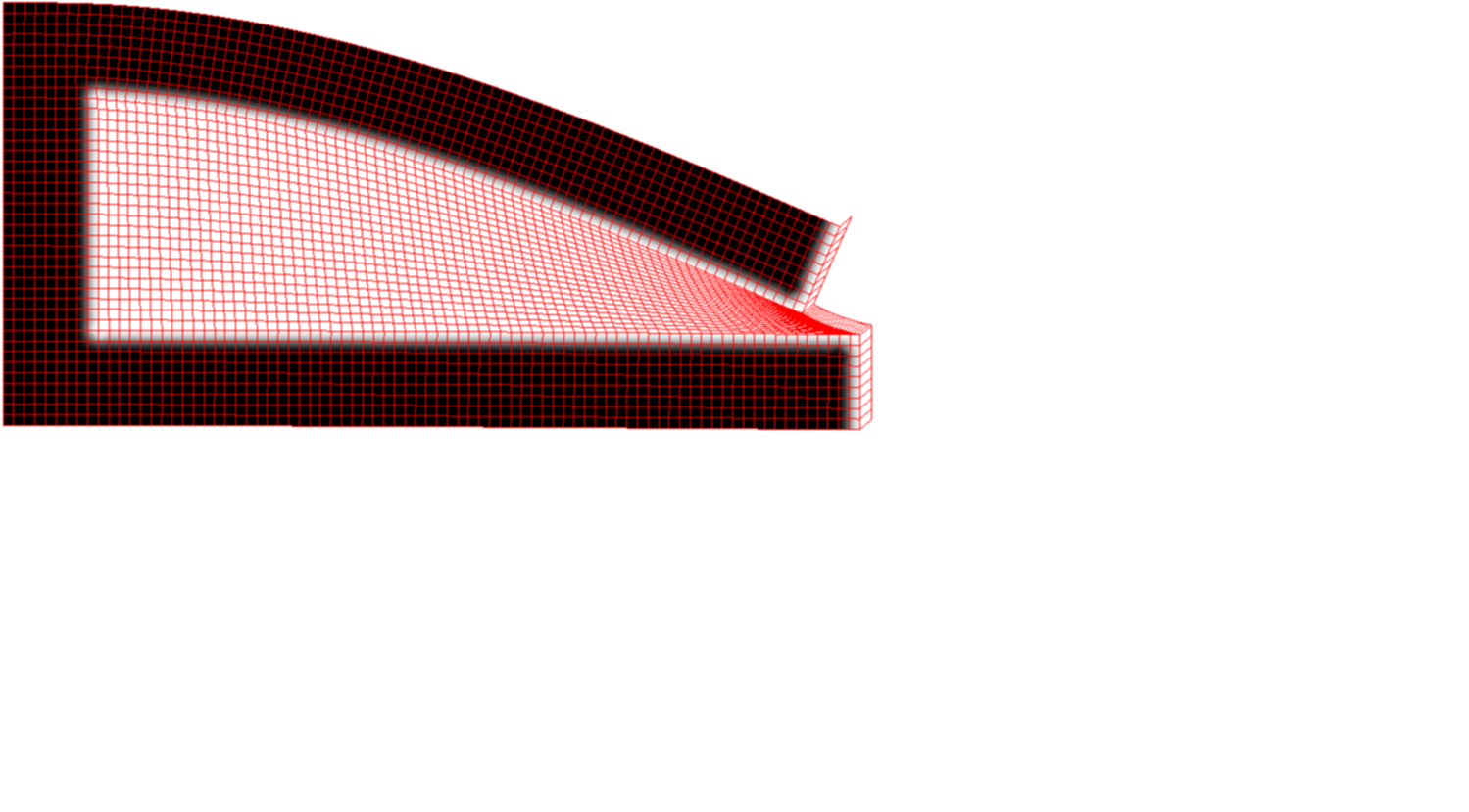}}\\[-5pt]
\subfloat[$u_{\scriptD y} = -H$]
{\includegraphics[scale=0.231, trim={0 150 590 0}, clip]
                 {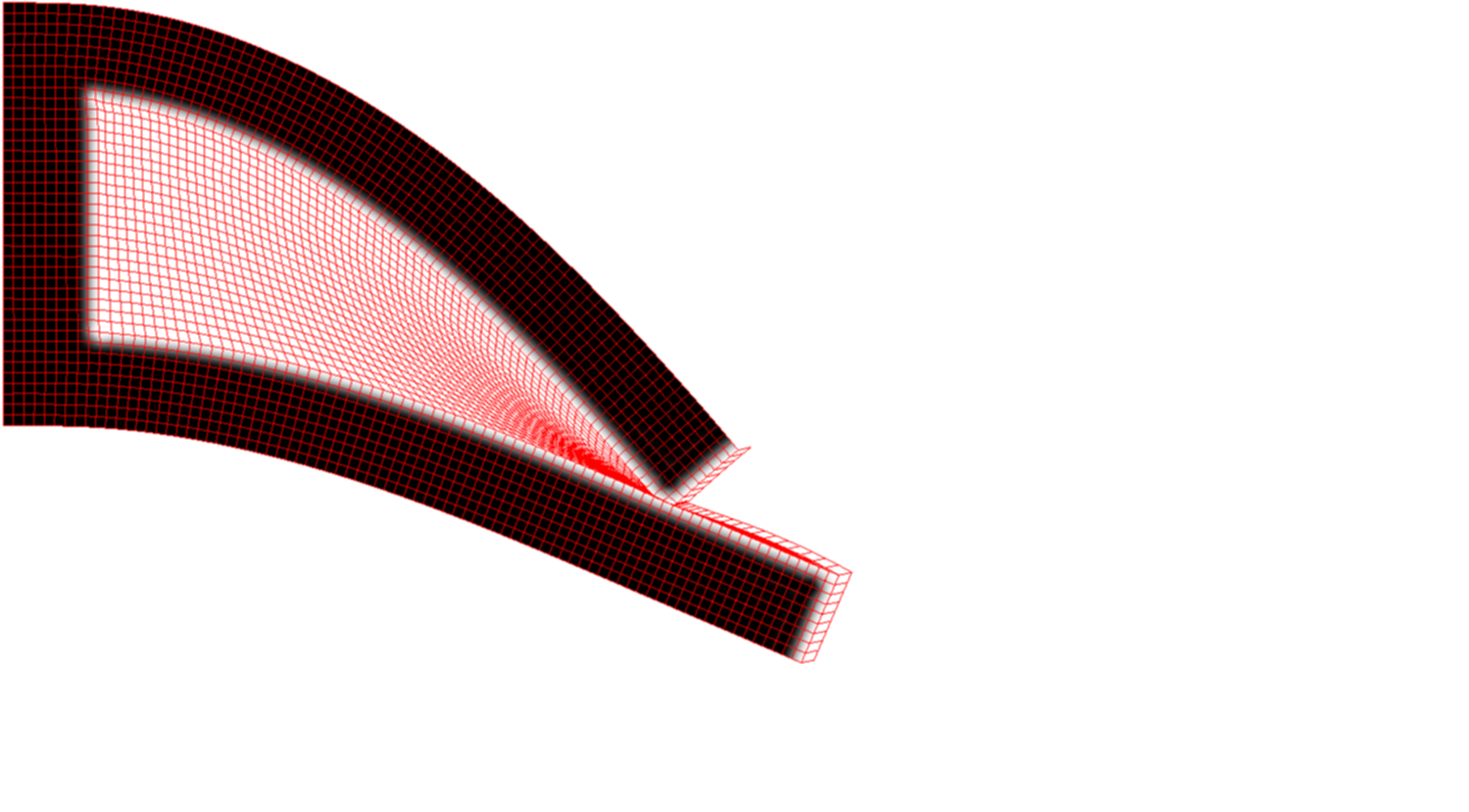}}
\caption{C-shape structure with $H\!=\!0.5L$ and graded interface in its underfomed state (a) and two deformed states at different load levels (b,c).}
\label{fig:double_c_contact_cont}
\end{figure}

Introducing a graded transition from void to solid will unavoidably affect the structure's mechanical behavior compared to a discontinuous transition, especially when studying the contact between solid members.
In order to evaluate the effectiveness of the third medium contact approach with the proposed void regularization, in the context of a graded void-solid interface, the last example from the previous subsection is reevaluated.
The domain discretization is now refined by a factor of two, resulting in $82\!\times\!40$ elements, and the transition from solid to void, i.e. from $\designvar\!=\!-10$ to $\designvar\!=\!10$, is spread over two elements, with the transition being linear in terms of $\designvar$.
In order to match the previous example, the minimum stiffness factor in Eq.~\eqref{eq:ramp} is set to $\mathcal{E}_0\!=\!10^{-12}$.
Figure~\ref{fig:double_c_contact_cont} shows the resulting material density field and the deformed structure at different loading states.

\begin{figure}[t!]
    \centering
    \includegraphics[width=84mm]{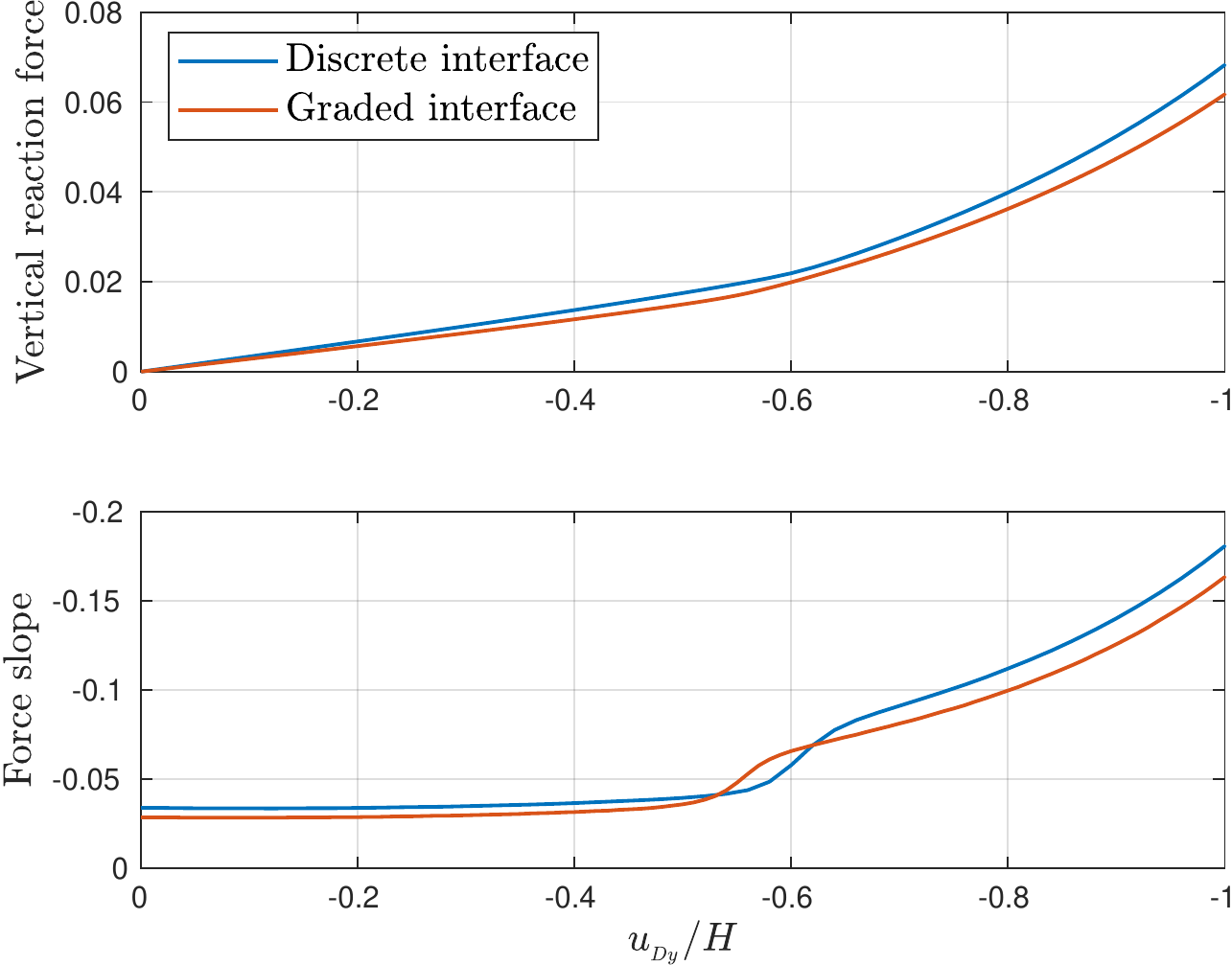}
    \caption{Load and stiffness response as function of prescribed displacement for the discontinuous void-solid transition (Figure~\ref{fig:double_c_contact}) and the graded void-solid interface (Figure~\ref{fig:double_c_contact_cont}).}
    \label{fig:cont_rho_load_dist}
\end{figure}

Figure~\ref{fig:cont_rho_load_dist} provides a more quantitative comparison between the model with the discontinuous solid to void transition from Figure~\ref{fig:double_c_contact} and the model with the graded transition from Figure~\ref{fig:double_c_contact_cont}.
The upper diagram of Figure~\ref{fig:cont_rho_load_dist} shows the reaction force as a function of the imposed vertical displacement to the upper beam for the two investigated cases, while the lower diagram shows the corresponding curve slopes.

The observed deviations between the two cases are due to different effects.
The interpretation of the original discontinuous interface as the graded interface shown in Figure~\ref{fig:double_c_contact_cont}a, in combination with the RAMP interpolation applied to the intermediate density material, leads to a reduced effective beam thickness due to the lower stiffness of the intermediate density material at the interface.
For this reason, the graded interface model is in general more compliant than the one with the discontinuous interface.
It is an accepted effect in topology optimization, that designs with intermediate densities are in general more compliant than crisp solid and void designs.

Apart from that, additional effects emerge in the context of contact modeling.
Most significantly due to the finite thickness of the graded interface between solid and void, contact now occurs prematurely as seen in Figure~\ref{fig:double_c_contact_cont}c and the lower diagram of Figure~\ref{fig:cont_rho_load_dist}.
Moreover, the presence of a layer of intermediate density material between the contacting solids adds some compliance to the contact, resulting in a less sharp transition from non-contact to contact, as observed e.g. in the force displacement curves shown.

Despite the presence of the mentioned effects, it is to be concluded that the overall contact behavior is captured adequately for topology optimization applications.
Especially the inaccuracy of remote interaction of solid members, can actually be an advantage in the context of optimization, because it provides information about potential contact interfaces in the surroundings that should be avoided or exploited, depending on the goal of the optimization and makes the contact problem differentiable.

\FloatBarrier
\section{Topology optimization formulation}
The method for third medium contact with the void regularization presented above, can in theory be implemented in every topology optimization framework for finite strains.
This section presents its implementation in a new generalized framework, which is based on the design parametrization from subsection~\ref{subsec:material_interpolation}, and a discretization-agnostic formulation.
The proposed optimization scheme is entirely defined in the continuous setting in weak form and is second order consistent.

\subsection{Optimization framework}
Consider an optimization problem defined as the minimization of an objective function $\cc$, i.e.
\begin{equation}
  \min_{\designvar}~\cc(\designvar,u,\multu)
\label{eq:opt_problem}
\end{equation}
subjected to the mechanical equilibrium constraint
\begin{equation}
  \RR(\designvar,u,\multu;\,\delta u,\delta\multu)=0~~~~\forall \delta u,\delta\multu
\label{eq:opt_constraint}
\end{equation}
corresponding to Eq.~\eqref{eq:equilibrium} but with the residual function $\RR$ also including the dependence on the design variable $\designvar$ introduced in Eq.~\eqref{eq:equilibriumresidual2}.

A simple but still rather general form for the objective function $\cc$ is the following additive split in three single variable functions
\begin{equation}
    \cc(\designvar, u, \multu) = \cc_\designvar(\designvar) 
    + \cc_u(u) 
    + \cc_\multu(\multu).
\label{eq:objsplit}
\end{equation}
Minimization of the objective function under the afore\-stated mechanical equilibrium constraint, is equivalent to minimization of the augmented objective function
\begin{equation}
\cc^*(\designvar,u,\multu) = \cc(\designvar,u,\multu)+\RR(\designvar,u,\multu;\,\ADJu,\ADJmultu)\\
\label{eq:obj_adjoint}
\end{equation}
for any $\ADJu$ and $\ADJmultu$ respectively in the spaces of $\delta u$ and $\delta q$.
The optimality condition is based on the variation of the objective function $\cc$ which can be evaluated as
\begin{equation}
\begin{split}
\delta \cc
=&~\delta \cc^*\\
=&~~\cc_{\designvar,\designvar}(\designvar;\,\delta \designvar)
  +\RR_{,\designvar}(\designvar,u,\multu;\,\ADJu,\ADJmultu; \delta \designvar)\\
 &+\cc_{u,u}(u;\,\delta u)
  +\RR_{,u}(\designvar,u,\multu;\,\ADJu,\ADJmultu; \delta u)\\
 &+\cc_{\multu,\multu}(\multu;\,\delta\multu)
  +\RR_{,\multu}(\designvar,u,\multu;\,\ADJu,\ADJmultu; \delta \multu).
\end{split}
\label{eq:varobj}
\end{equation}
Applying the adjoint sensitivity analysis method, the variations $\delta u$ and $\delta\multu$ are eliminated from Eq.~\eqref{eq:varobj} by finding multipliers $\ADJu$ and $\ADJmultu$ that satisfy the adjoint equations
\begin{equation}
\begin{split}
 &\RR_{,u}(\designvar,u,\multu;\,\ADJu,\ADJmultu; \delta u)
  + \cc_{u,u}(u;\,\delta u)
  = 0 ~~\forall~\delta u \\
 &\RR_{,\multu}(\designvar,u,\multu;\,\ADJu,\ADJmultu; \delta\multu)
  + \cc_{\multu,\multu}(\multu;\,\delta\multu)
  = 0 ~~\forall~\delta\multu.
\end{split}
\label{eq:adjeqn}
\end{equation}
The variation of the objective is thereby reduced to
\begin{equation}
\delta \cc
= \cc_{\designvar,\designvar}(\designvar;\,\delta \designvar)
  +\RR_{,\designvar}(\designvar,u,\multu;\,\ADJu,\ADJmultu; \delta \designvar).
\label{eq:varobj_short}
\end{equation}

In total, mechanical equilibrium, adjoint analysis and optimality are expressed by the following system of coupled nonlinear equations
\begin{equation}
\begin{split}
   &\RR(\designvar,u,\multu;\,\delta u,\delta\multu)=0~~~~\forall ~\delta u,\delta\multu\\
   &\RR_{,u}(\designvar,u,\multu;\,\ADJu,\ADJmultu; \delta u)
    + \cc_{u,u}(u;\,\delta u)
    = 0 ~~\forall~\delta u \\
   &\RR_{,\multu}(\designvar,u,\multu;\,\ADJu,\ADJmultu; \delta\multu)
   + \cc_{\multu,\multu}(\multu;\,\delta\multu)
   = 0 ~~\forall~\delta\multu\\
   &\cc_{\designvar,\designvar}(\designvar;\,\delta \designvar)
  +\RR_{,\designvar}(\designvar,u,\multu;\,\ADJu,\ADJmultu; \delta \designvar)=0 ~~\forall~\delta\designvar
\end{split}
\label{eq:opt_system}
\end{equation}
with regard to the unknowns $\designvar$, $u$, $\multu$, $\ADJu$ and $\ADJmultu$.
Assuming that all functions involved in Eq.~\eqref{eq:opt_system} are differentiable with respect to the five unknowns, a consistent linearization of this system of equations is trivial to obtain in order to apply Newton's method.
Application of Newton's method to the fully coupled Eq.~\eqref{eq:opt_system} corresponds to a second order optimization scheme and it will lead to quadratic converge, simultaneously towards the optimal design and mechanical equilibrium, if the initial guess is sufficiently close to the solution.

Due to the extremely nonlinear nature of the design problem though, the initial guess for $\designvar$ will never be close enough to the final design for Newton's algorithm to converge.
A reasonable strategy for dealing with this without losing the second order information on the system is to add a damping on the last equation in Eq.~\eqref{eq:opt_system} and solve the system as an ODE with an implicit time discretization, where each time step is solved with Newton's method.
Somewhat similar to reference \cite{2009Kl-To}, but also introducing damping of spatial gradients of $\designvar$, the optimality condition $\delta \cc = 0$, appearing in the last row of Eq.~\eqref{eq:opt_system}, is now substituted with
\begin{equation}
\delta \cc + 
\int_\Omega
\Big\{\dot{\designvar} \delta \designvar
+l_t^2\,\nabla \dot{\designvar}\cdot\nabla \delta \designvar\Big\}
~d\Omega= 0~~\forall~\delta\designvar
\label{eq:design_opt_damped}
\end{equation}
where time derivatives are with regard to a pseudo-time $t$, and $l_t$ is a characteristic length-scale for the spatial evolution of the design field $\designvar$.
Assuming that the design field $\designvarold$ at pseudo-time $t-\Delta t$ is known, the design field $\designvar$ at pseudo-time $t$ can be found by solving the system
\begin{equation}
\begin{split}
   &\RR(\designvar,u,\multu;\,\delta u,\delta\multu)=0~~~~\forall ~\delta u,\delta\multu\\
   &\RR_{,u}(\designvar,u,\multu;\,\ADJu,\ADJmultu; \delta u)
    + \cc_{u,u}(u;\,\delta u)
    = 0 ~~\forall~\delta u \\
   &\RR_{,\multu}(\designvar,u,\multu;\,\ADJu,\ADJmultu; \delta\multu)
   + \cc_{\multu,\multu}(\multu;\,\delta\multu)
   = 0 ~~\forall~\delta\multu\\
   &\cc_{\designvar,\designvar}(\designvar;\,\delta \designvar)
  +\RR_{,\designvar}(\designvar,u,\multu;\,\ADJu,\ADJmultu; \delta \designvar)\\
  &+\int_\Omega\Bigg\{
  \dfrac{\designvar - \designvarold}{\Delta t} \delta \designvar\\
  &~~~~~~~~~~+l_t^2 \dfrac{\nabla \designvar - \nabla \designvarold}{\Delta t}\cdot\nabla \delta \designvar
  \Bigg\}~d\Omega = 0~~\forall~\delta\designvar
\end{split}
\label{eq:opt_system_damped}
\end{equation}
obtained from substituting a backward Euler approximated version of Eq.~\eqref{eq:design_opt_damped} into Eq.~\eqref{eq:opt_system}.

It is easy to verify that the steady state solution of Eq.~\eqref{eq:opt_system_damped} corresponds to an exact solution of Eq.~\eqref{eq:opt_system}.
Hence, the introduced damping constitutes a consistent regularization in the sense that the exact solution is recovered as the time step $\Delta t$ approaches infinity.
At the same time, the convex term added to the original optimality equation ensures that, for $\Delta t$ small enough, the last row of Eq.~\eqref{eq:opt_system_damped} will not lead to a diverging Newton loop.
In that sense, Eq.~\eqref{eq:opt_system_damped} provides a very convenient way to start the optimization with small pseudo-time increments $\Delta t$, that ensure a stable design evolution, and end it with exceedingly large time steps $\Delta t$, that recover the quadratic convergence rate of the original Eq.~\eqref{eq:opt_system} towards its exact optimality solution.

\FloatBarrier
\subsubsection*{Solution algorithm}

\begin{figure}
\centering
\tikzstyle{block}    = [rectangle, rounded corners, fill=black!20,
                        minimum width=1cm, minimum height=0.5cm,
                        text centered, draw]
\tikzstyle{decision}  = [diamond, aspect=5,
                         minimum width=2cm, minimum height=0.1cm,
                         text centered, draw, fill=black!20]
\tikzstyle{line}      = [draw, -latex']

\begin{center}
\begin{tikzpicture}[node distance = 1cm]
    \node [block, text width=5.4cm, inner ysep=4pt] (init) 
    	{\small Intialize variables
         $\designvar$, $u$, $q$, $\lambda_u$ and $\lambda_q$\\
         and data $\designvarold$, $\Delta t$. Set $i = 0$.};
    \node [block, below of=init, node distance=1.1cm, text width=4cm, inner ysep=4pt]  (step)
    	{\textbf{\small Solve design update i}};
    \node [block, below of=step, node distance=0.9cm, text width=4cm, inner ysep=4pt]  (NR)
    	{\small Newton solve of Eq.~\eqref{eq:opt_system_damped}};
    \node [decision, below of=NR, node distance=1.4cm, text width=3.1cm, inner sep=0pt, aspect=3]  (conv)
    	{\small Newton converged within 7 iterations?};
    \node [block, right of=conv, node distance=3.6cm] (delta_t)
    	{\small $\Delta t \leftarrow \Delta t/4$};
    \node [block, below of=conv, node distance=1.9cm, text width=4cm] (update)
    	{\small
    	 $\designvarold\leftarrow \designvar$\\
    	 $t\leftarrow t\!+\!\Delta t$\\
    	     	 If Newton iterations $\leq$ 5: \\ $\Delta t \leftarrow 2\Delta t$
    	 };
    \node [decision, below of=update, node distance=1.5cm, text width=2cm, inner sep=0pt, aspect=4]  (itermax)
    	{\small $i=i_{\max}$ or \\ $\Delta t > \Delta t_{\max}$?};
    \node [block, right of=itermax, node distance=3.5cm, text width=1.4cm] (next)
    	{\small $i \leftarrow i\!+\!1$};
    \node [block, below of=itermax, node distance=1.25cm, text width=3.5cm] (final)
    	{\small Optimization completed};

  \path [line] (init) -- (step);
  \path [line] (step) -- (NR);
  \path [line] (NR) -- (conv);
  \path [line] (conv) -- (update);
  \path [line] (conv) -- (delta_t);
  \path [line] (delta_t.north) |-  (NR.east);
  \path [line] (update) -- (itermax);
  \path [line] (itermax) -- (next);
  \path [line] (itermax) -- (final);
  \path [line] (next.east) -| (4.7,-2) |-  (step.east);

  \node (conv_y)  at ([shift={(0.3,-0.9)}]conv){\footnotesize yes};
  \node (conv_n)  at ([shift={(2.45,0.15)}]conv){\footnotesize no};
  
  \node (iter_y)  at ([shift={(0.3,-0.75)}]itermax){\footnotesize yes};
  \node (iter_n)  at ([shift={(2.4,0.15)}]itermax){\footnotesize no};
  \end{tikzpicture}
\end{center}
\caption{Flow chart of the adaptive time step solution algorithm.}
\label{fig:sol_alg}
\end{figure}
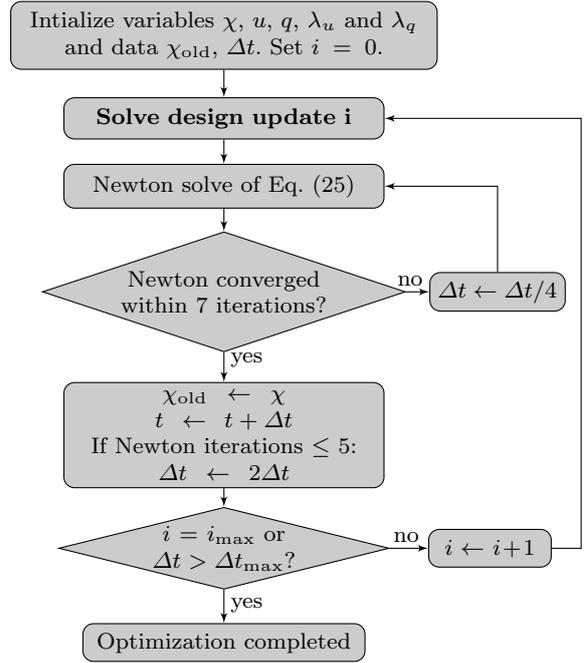
Due to its importance for the overall efficiency of the proposed topology optimization framework, the exact adaptive time stepping algorithm for Eq.~\eqref{eq:opt_system_damped} is documented schematically in Figure~\ref{fig:sol_alg}.
The basic idea is to use the convergence behavior of the Newton loop on Eq.~\eqref{eq:opt_system_damped} as a measure of the nonlinearity of the system and adapt $\Delta t$ accordingly.
In practice this leads to small time steps when rapid design changes occur and large time steps when design changes are small. 

\subsection{Application to load-displacement path control}
In order to demonstrate the capabilities of the proposed contact modeling approach in topology optimization, the next section deals with a load-displacement path optimization problem.
For this reason, the general framework introduced in the previous subsection needs to be applied to this specific optimization problem.

Assuming a set of prescribed displacements $u_{\scriptD}^{\scriptj}$ for $j\!=\!1,2,...$ in the mechanical residual Eq.~\eqref{eq:equilibriumresidual}, the goal is to find a design field $\designvar$ that minimizes the deviation between the respective reaction tractions $\multu^{\scriptj}$ obtained, and a set of corresponding prescribed average reactions $\multu^{\scriptj}_*$.
This can be achieved by minimizing the squared deviation between the actual and desired tractions through an objective function in the form
\begin{equation}
    \cc_\multu(\multu) = \sum_{j} \int_{\Gamma_{\scriptD}} 
    w^{\scriptj} \norm{\multu^{\scriptj} - \multu^{\scriptj}_* }^2~d\Gamma,
    \label{eq:obj_multu}
\end{equation}
where $w^{\scriptj}$ are individual weighting factors for each control point.
To maintain the compact notation of the previous subsection, all variables $\multu^{\scriptj}$ are combined in a set $\multu\!=\!\{\multu^{\left<1\right>},\multu^{\left<2\right>},...\}$.
The corresponding set of displacement fields $u^{\scriptj}$ are combined in a set $u\!=\!\{u^{\left<1\right>},u^{\left<2\right>},...\}$.

With this notation, the first row of Eq.~\eqref{eq:opt_system_damped} can express the mechanical equilibrium at all control points $j\!=\!1,2,...$, which are otherwise assumed to be independent from each other, i.e. excluding path dependent mechanical responses.
Corresponding adjoint problems need to be defined for each control point $j$.
In order to express these in the second and third row of Eq.~\eqref{eq:opt_system_damped}, sets of the involved adjoint variables are defined as $\ADJu\!=\!\{\ADJu^{\left<1\right>},\ADJu^{\left<2\right>},...\}$ and $\ADJmultu\!=\!\{\ADJmultu^{\left<1\right>},\ADJmultu^{\left<2\right>},...\}$.

So far, there has been no dependence between mechanical equilibrium or adjoint problems for different control points~$j$.
This is not true anymore when considering the damped optimality equation in the last row of Eq.~\eqref{eq:opt_system_damped}, where the term
\begin{equation}
\begin{split}
    &\RR_{,\designvar}(\designvar,u,\multu;\,\ADJu,\ADJmultu; \delta \designvar)\\
    &~~=\sum_j \RR_{,\designvar}(\designvar,u^{\scriptj},\multu^{\scriptj};\,\ADJu^{\scriptj},\ADJmultu^{\scriptj}; \delta \designvar)
\end{split}
\label{eq:adjoint_contrib_multipoint}
\end{equation}
couples all subproblems for the various control points involved in the load-displacement path optimization.
As a consequence, in order to maintain a second order optimization scheme, Eq.~\eqref{eq:opt_system_damped} needs to be solved as a coupled system of equations involving all variables at the different control points simultaneously.

Regarding the contribution $\cc_u(u)$ in the objective function in Eq.~\eqref{eq:objsplit}, there is no need for this kind of contribution in this specific application.
Hence
\begin{equation}
    \cc_u(u) = 0.
    \label{eq:obj_u}
\end{equation}

On the contrary, the design field contribution $\cc_\designvar(\designvar)$ in Eq.~\eqref{eq:objsplit} is essential both
for controlling the width of the graded void-solid interface and for enforcing material utilization constraints.
By observing Figure~\ref{fig:projection}, the width of the void-solid interface for a linearly varying $\designvar$ from large negative to large positive values is approximately equal to $2/(\rho'(0) \norm{\nabla \designvar})$.
Enforcing a minimum width of the diffuse interface equal to $l_i$ can therefore be obtained by penalizing positive values of the quantity
\begin{equation}
    \Big< \rho'(0) \norm{\nabla \designvar}-2/l_i\Big>,
    \label{eq:lengthscale}
\end{equation}
for example by means of a p-norm over the entire design domain.
This kind of penalization is essential not only for providing a length-scale for the void-solid interface but also for obtaining a converged design from Eq.~\eqref{eq:opt_system_damped}, with the otherwise unbounded design variable $\designvar$.

Based on a dimensionless version of Eq.~\eqref{eq:lengthscale} as well as on an enforced upper bound $\bar{\rho}_{\max}$ to the average density $\bar{\rho}$ over the design domain, i.e. $\bar{\rho} \le \bar{\rho}_{\max}$, an extended function $\cc_\designvar$ is defined as
\begin{equation}
\begin{split}
   &\cc_\designvar(\designvar,\bar{\rho},\mu_\mathrm{avg},\mu_{\max})\\
   &~~
   \begin{split}
     = \int_\Omega& \bigg\{
     \kint \Big< \rho'(0) \norm{\nabla \designvar}l_i/2-1\Big>^n \\
     &+ \left( \rho(\designvar) - \bar{\rho} \right) \mu_\mathrm{avg}\\
     &+ \dfrac{\kvol}{2} \left< \frac{\bar{\rho}}{\bar{\rho}_{\max}} -1 -\frac{\mu_{\max}}{\kvol} \right>^2
     -\frac{\mu_{\max}^2}{2 \kvol}
     \bigg\}~d\Omega.
   \end{split}
\end{split}
\end{equation}
In this expression, $\kint$ is a weighting factor and $n$ is a p-norm exponent for the enforcement of the minimum interface width $l_i$.
The equality of the scalar variable $\bar{\rho}$ to the average of the field $\rho$ over $\Omega$ is enforced through the Lagrangian with the Lagrange multiplier $\mu_\mathrm{avg}$.
The inequality constraint $\bar{\rho} \le \bar{\rho}_{\max}$ is enforced through the augmented Lagrangian with the Lagrange multiplier $\mu_{\max}$ and the augmentation constant $\kvol$.
The additional equations necessary for solving for the three new scalar unknowns $\bar{\rho}$, $\mu_\mathrm{avg}$ and $\mu_{\max}$ are provided by saddle point conditions of the respective Lagrangians, i.e.
\begin{equation}
\begin{split}
    &\cc_{\designvar,\bar{\rho}}(\bar{\rho},\mu_\mathrm{avg},\mu_{\max}) = 0\\
    &\cc_{\designvar,\mu_\mathrm{avg}}(\designvar,\bar{\rho}) = 0\\
    &\cc_{\designvar,\mu_{\max}}(\bar{\rho},\mu_{\max}) = 0
\end{split}
\label{eq:saddle_points}
\end{equation}
These three equations need to be appended to Eq.~\eqref{eq:opt_system_damped}, linearized and solved simultaneously with all other problem variables.

With the adaptations presented in the present subsection so far, the application of the framework to load-displacement path control is essentially completed.
Nevertheless, for more clarity, the most complex terms involved in the solution of Eq.~\eqref{eq:opt_system_damped} are presented in more detail.
With $\cc_{u,u}$ vanishing due to Eq.~\eqref{eq:obj_u}, the adjoint equation in the second row of Eq.~\eqref{eq:opt_system_damped} actually reads
\begin{equation}
\begin{split}
&
\begin{split}
    \int_\Omega 
    &\bigg\{
    \ramp{\designvar}\left(\diff{P_\solid^{\scriptj}}{\nabla u^{\scriptj}}:\nabla\delta u\right) : \nabla\ADJu^{\scriptj}\\
    &~+ \kreg\lock{\designvar}e^{-5 \det{F_j}}\\
    &~~~\Big(\!\Hessian{\delta u}-5 \det{F_j}\!\left(F_j^{-1}\!:\!\nabla\delta u\right) \Hessian{u^{\scriptj}}\!\Big)
    \!\tensorm\!\Hessian{\ADJu^{\scriptj}}\!\bigg\}\,d\Omega\\
\end{split}\\
&+\int_{\Gamma_{\scriptD}} \delta u \cdot \ADJmultu^{\scriptj}~d\Gamma= 0~~~\forall~j,\delta u
\end{split}
\label{eq:adju}
\end{equation}
where $F_j=F(\nabla u^{\scriptj})$.

It should be noted that applying Newton's method on Eq.~\eqref{eq:opt_system_damped} requires linearization of Eq.~\eqref{eq:adju} with respect to all involved unknowns, which means that the derivative of the 1\textsuperscript{st} Piola-Kirchhoff stress $P$ with respect to $\nabla u$, appearing in this equation, needs to be differentiated one additional time with respect to $\nabla u$.
This is easy for the adopted hyperelastic material law with the resulting expression for $P$ shown in Eq.~\eqref{eq:PK1}, but this is not in general the case for any constitutive law.
The term in Eq.~\eqref{eq:adju} due to the proposed void regularization, weighted with $\kreg$, is also straightforward to linearize with respect to $\nabla u$.

For this specific application, the adjoint equation in the third row of Eq.~\eqref{eq:opt_system_damped} becomes
\begin{equation}
\begin{split}
    \int_{\Gamma_{\scriptD}}\Big\{&
    \delta\multu \cdot \ADJu^{\scriptj}\\
    &+ 2 w^{\scriptj}\!\left( \multu^{\scriptj} - \multu^{\scriptj}_* \right) \cdot \delta\multu
    \Big\}~d\Gamma = 0 ~~\forall ~j,\delta\multu
\end{split}
\label{eq:adjmultu}
\end{equation}

Finally, the damped optimality equation for a design update from a previous design $\designvarold$, appearing last in Eq.~\eqref{eq:opt_system_damped}, is particularized for this application as
\begin{equation}
\begin{split}
\int_\Omega\Bigg\{
    &n\bigg<\rho'(0) \norm{\nabla \designvar}\dfrac{l_i}{2}-1\bigg>^{n-1}\!
      \rho'(0)\dfrac{\nabla \designvar}{\norm{\nabla \designvar}}\cdot{\nabla\delta \designvar}\\
    &\!+ \rho'(\designvar) \mu_\mathrm{avg} ~\delta \designvar \\
    &
    \begin{split}
       +&\sum_j
       \Big(\rampder{\designvar}\,P_\solid(\nabla u^{\scriptj})\!:\!\nabla\ADJu^{\scriptj}\\
       &~~+\kreg\lockder{\designvar}\,e^{-5\det{F\left(\nabla u^{\scriptj}\right)}}\Hessian{u^{\scriptj}}\!\tensorm\!\Hessian{\ADJu^{\scriptj}}
       \Big)\delta \designvar
    \end{split}\\
    &\!+\dfrac{\designvar\!-\!\designvarold}{\Delta t} \delta \designvar\\
    &\!+l_t^2 \dfrac{\nabla \designvar\!-\!\nabla \designvarold}{\Delta t}\cdot\nabla \delta \designvar
\Bigg\}~d\Omega= 0~~~\forall~~\delta\designvar.
\end{split}
\label{eq:design_update}
\end{equation}

Despite some lengthy expressions, all involved terms are differentiable also in this last equation, so that application of Newton's method with a consistent tangent matrix is possible.
The model is implemented in the automated finite element framework GetFEM \cite{2021Re-Po}, which performs necessary symbolic linearizations and the numerical assembly of the discretized system of equations.

\FloatBarrier
\section{Results and discussion}
\begin{figure}[b]
    \centering
    \resizebox{84mm}{!}{\input{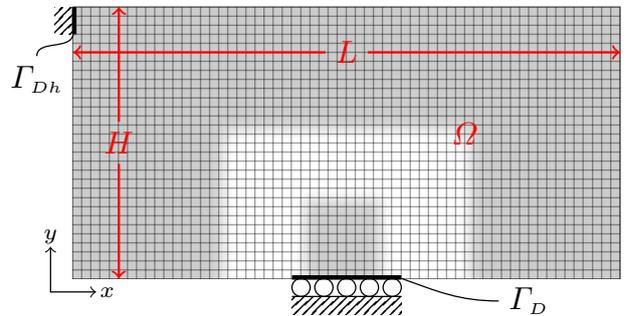}}
    \caption{Problem domain for designing a stiff coupling with initial clearance. Homogeneous Dirichlet condition region $\Gamma_{\scriptD h}$ has a length 0.1H and the nonhomogeneous Dirichlet condition region $\Gamma_{\scriptD}$ spans from $0.4L$ to $0.6L$ along the bottom edge. Dimensions: $L = 52$, $H = 26$.
    }
    \label{fig:ex1_domain}
\end{figure}

The internal contact enabled topology optimization formulation for load-displacement path control, described in the previous section, is applied now to design a structure that switches from a completely decoupled to a high stiffness connection, within a given displacement interval.
Figure~\ref{fig:ex1_domain} shows the problem domain, along with the initial material density distribution and a discretization with $60\!\times\!30$ square elements.
A small region $\Gamma_{\scriptD h}$, near the top of the left side of the domain, is fixed, while a prescribed horizontal displacement is imposed on a region $\Gamma_{\scriptD}$ in the middle of the bottom side of the domain.
The path control problem to be solved consists in minimizing the reaction force on $\Gamma_{\scriptD}$ for a horizontal displacement $u_{\scriptD x}^{\scriptnum{1}}\!=\!0.1 L$ and at the same time maximize the reaction force for $u_{\scriptD x}^{\scriptnum{2}}\!=\!0.15 L$.

Before applying the method, a suitable discretization needs to be specified for all unknown fields.
The displacement fields $u^{\scriptj}$ are approximated with 8-node quadratic elements.
All degrees of freedom within the $\Gamma_{\scriptD h}$ region and all vertical degrees of freedom within the $\Gamma_{\scriptD}$ region are removed from the finite element space.
Moreover, all horizontal degrees of freedom within $\Gamma_{\scriptD}$ are reduced to a single master degree of freedom, enforcing a rigid body translation of this boundary.
The finite element space resulting after these restrictions is used for $u^{\scriptj}$, the corresponding adjoint fields $\ADJu^{\scriptj}$ and variations $\delta u$.
Linear 4-node elements are used for approximating the design field $\designvar$ and the corresponding variations $\delta\designvar$ without any further restrictions.
Since $\chi$ is unbounded, it does not suffer by the usual limitation to linear elements when approximating the density field $\rho$ directly, in order to ensure density values in the $[0,1]$ interval.
Nevertheless, numerical experiments with quadratic elements for $\designvar$ showed an only relatively small accuracy gain.
For this choice of finite element spaces for the different fields, all terms involved in the the problem domain $\Omega$ are assembled with nine Gauss points in each quadrilateral element.
Terms defined on the boundary $\Gamma_{\scriptD}$ are assembled with the corresponding three Gauss points per element edge.

Since vertical displacements on $\Gamma_{\scriptD}$ are already fixed by restricting the relevant finite element space, the unknown multipliers $\multu^{\scriptj}$ on $\Gamma_{\scriptD}$ are reduced to just their $x$ component, written in the form $\multu^{\scriptj}\!=\!\{\multu^{\scriptj}_x,0\}^T$.
The same format applies to the corresponding adjoint variables $\ADJmultu^{\scriptj}$, variations $\delta\multu$, and target reactions $\multu^{\scriptj}_*$ , which also reduce to scalars.

\begin{table}[t]
\caption{Model parameters for the optimization example.}
\label{tab:example_params}
\setlength{\tabcolsep}{2pt}
\begin{tabular}{l c c c}\hline \\[-8pt]
\centering
Domain dimension & $L \times H$ & $52 \times 26$ & mm\textsuperscript{2}\\
Mesh size & & $60 \times 30$ & -\\[5pt]

Solid bulk modulus & $K_s$ &  $5/3$ & MPa\\ 
Solid shear modulus & $G_s$ & $5/14$ & MPa\\ 
Void/solid stiffness contrast & $\mathcal{E}_0$ & $10^{-12}$  & -\\
Void regularization scaling & $\kregdimless$ & $10^{-6}$ & -\\[5pt]

Max. material vol. fraction & $\bar{\rho}_{\max}$ &  0.35 & -\\
Vol. constr. augm. param.& $k_\rho$ &  $10^{-3}$ & -\\[5pt]

Control point weights & $w^{\scriptnum{1}}$/$w^{\scriptnum{2}}$ & $10^4/10^3$ & MPa\textsuperscript{-2}\\
Target tractions & $\multu^{\scriptnum{1}}_{*x}$/$\multu^{\scriptnum{2}}_{*x}$ & $0$ / $0.1$ & MPa\\[5pt]

Transient length scale & $l_t$ & 2 & mm\\
Minimum interface width & $l_i$ & 2 & mm\\
Interf. width p-norm weight & $\kint$ & $10^{-1}$ & -\\
Interf. width p-norm exponent & n & 6 & -\\ \hline
\end{tabular}
\end{table}

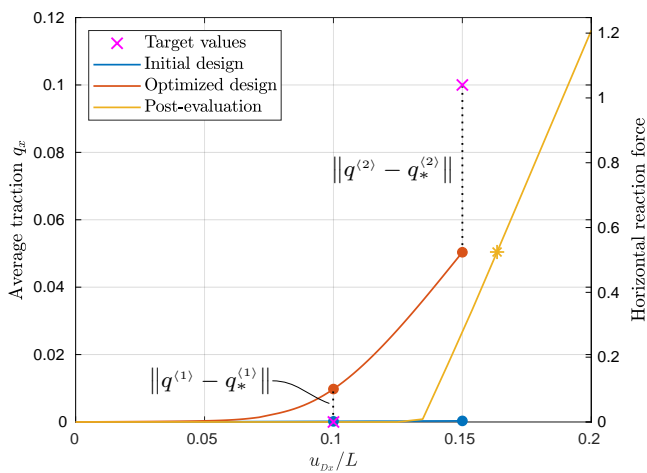
\begin{figure}[b]
    \centering
    \begin{tikzpicture}[every node/.style={anchor=south west,inner sep=0pt},x=1mm, y=1mm]
    \node (fig1) at (0,0) {\includegraphics[width=84mm]{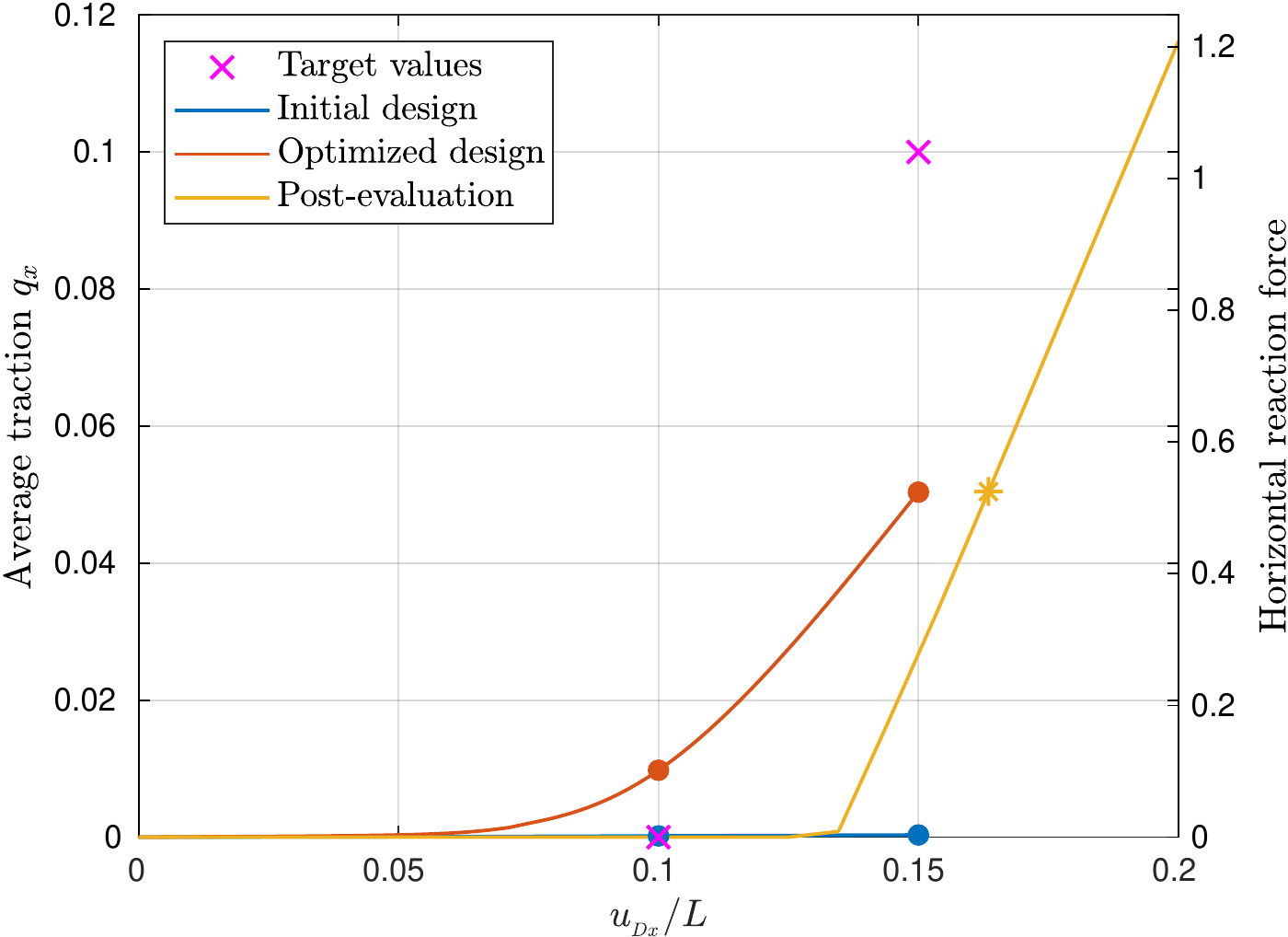}};
    \draw [dotted, line width=0.7] (60,30) -- (60,51);
    \node [anchor=east] at (59,40) {$\norm{q^{\scriptnum{2}} - q_{*}^{\scriptnum{2}}}$};
    \draw [dotted, line width=0.7] (43,7.5) -- (43,11);
    \node [anchor=east] (q1) at (35,12) {$\norm{q^{\scriptnum{1}} - q_{*}^{\scriptnum{1}}}$};
    \draw [line width=0.3] (q1.east) to[out=0, in=160] (42.5,9);
    \end{tikzpicture}
    \caption{Load-displacement response of the initial and optimized design, together with a post-evaluation analysis with a body fitted mesh. The colored dots mark traction values at the control points and the asterisk indicates the state shown in Figure~\ref{fig:ex1_posteval_deformed}.}
    \label{fig:ex1_load_disp}
\end{figure}

Based on this discretization, the optimization procedure from the previous section is applied with a volume constraint $\bar{\rho}_{\max}$ of 35\% and two control points $j\!=\!\{1,2\}$, illustrated in Figure~\ref{fig:ex1_load_disp}.
In order to minimize the initial stiffness of the coupling, a target average traction $\multu^{\scriptnum{1}}_{*x}\!=\!0$ is set for the first control point at $u_{\scriptD x}^{\scriptnum{1}}\!=\!0.1 L$.
Requiring a stiff coupling upon further displacement of the moving part is represented by a high target value $\multu^{\scriptnum{2}}_{*x} = 0.1$ for the second control point at $u_{\scriptD x}^{\scriptnum{2}}\!=\!0.15 L$.
The corresponding weightings $w^{\scriptj}$ for the objective function $\cc_\multu$ according to Eq.~\eqref{eq:obj_multu}, are provided in Table~\ref{tab:example_params}.
With this kind of objective $\cc_\multu$, a design is expected where contact onset occurs between the two control points.
All further parameters for this optimization example are reported in Table~\ref{tab:example_params}.

The initial design, shown in Figure~\ref{fig:ex1_domain}, consists of two disconnected regions with material density $\rho\!=\!0.2$, separated by a stripe of void material with $\rho\!=\!10^{-5}$.
Other non-uniform starting guesses like this one result in similar albeit shifted designs, whereas completely uniform initial designs lack of adequate initial non-linearity to guide the optimization towards a contact dominated structure.

\begin{figure}[b]
    \centering
    \includegraphics[width=84mm]{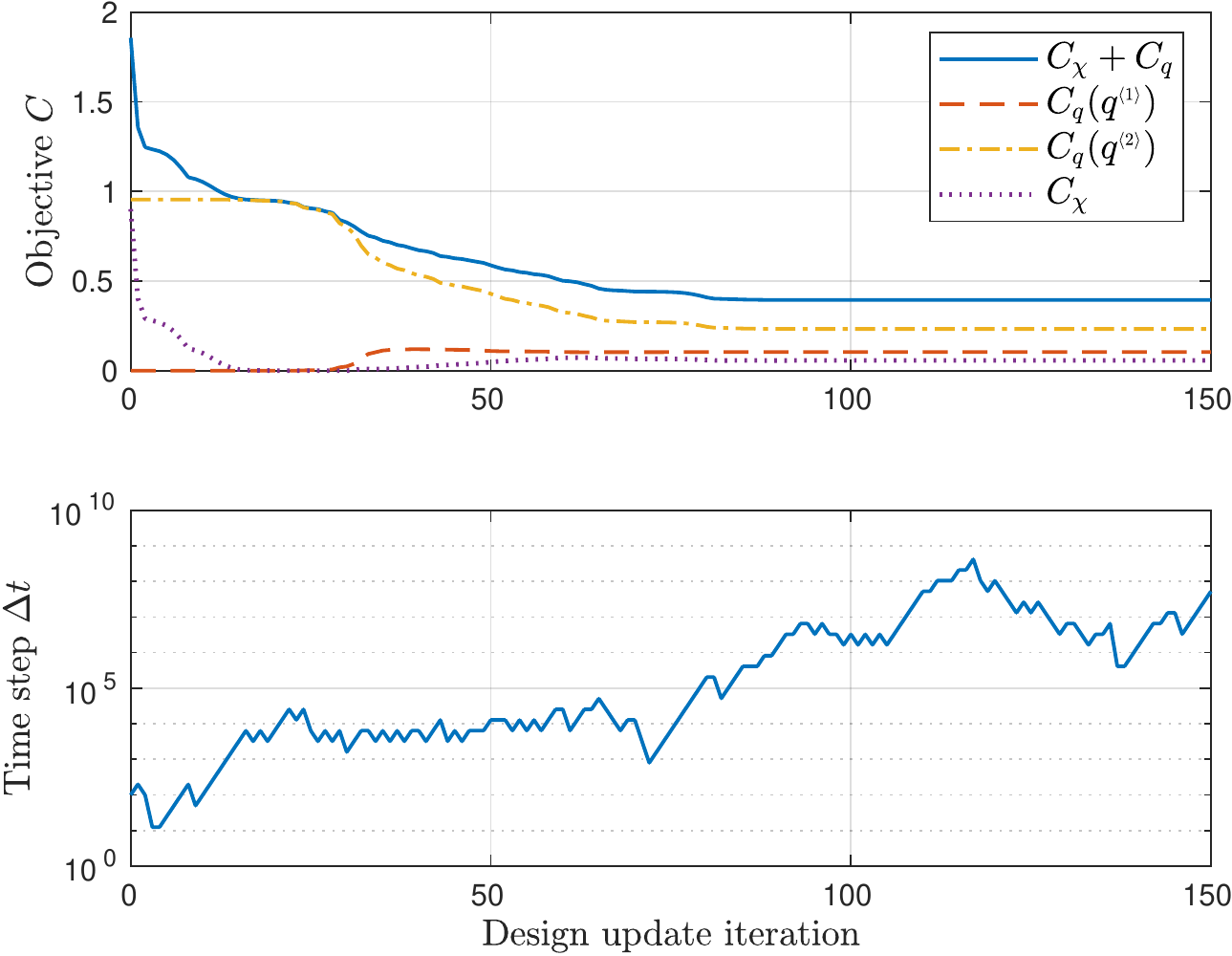}
    \caption{Evolution of objective value and design update time step. The total objective value $\cc\!=\!\cc_\designvar+\cc_\multu$ and the contributions of the two control points to $\cc_\multu$ are shown separately.}
    \label{fig:ex1_iter}
\end{figure}

\begin{figure}
\centering
    \begin{tikzpicture}[every node/.style={anchor=south west,inner sep=0pt},x=1mm, y=1mm]
        \node[anchor=south] at (0mm,0mm) {
            \includegraphics[scale=0.22,trim={6cm 22.5cm 6cm 0}, clip]
                            {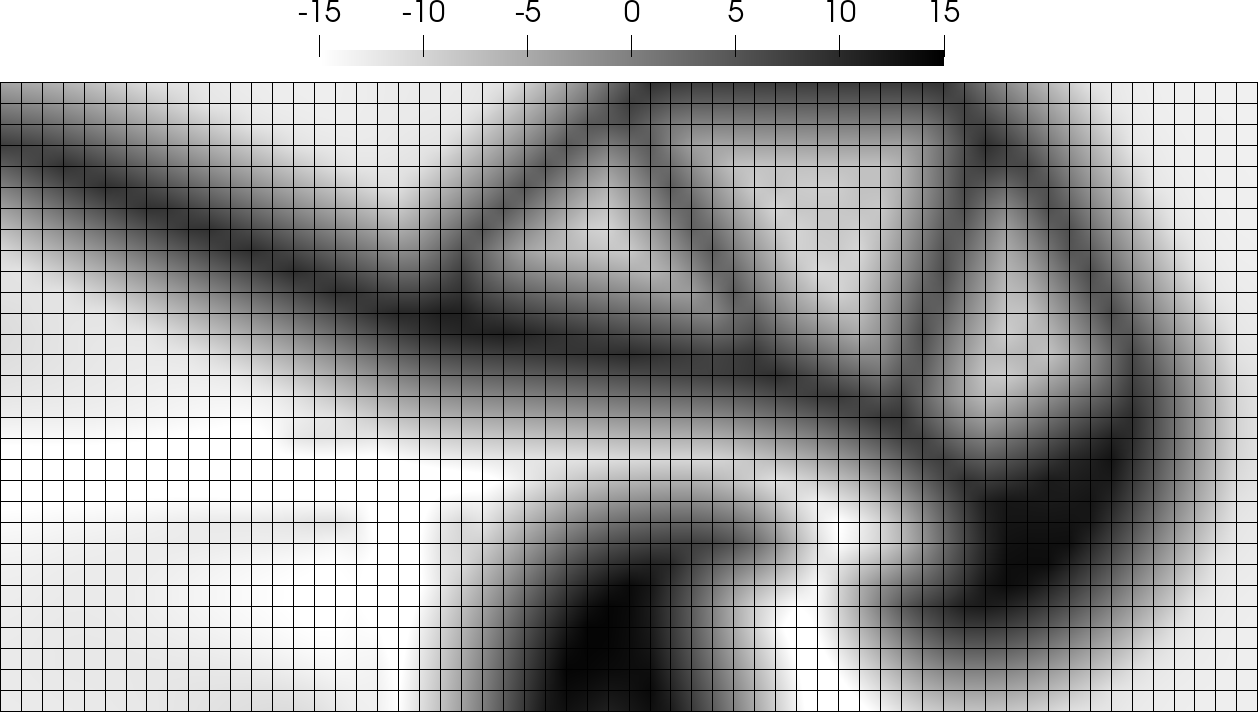}};
        \node[anchor=north] at (0mm,0mm) {
            \includegraphics[scale=0.17,trim={0 0 0 2.5cm}, clip]
                            {fig/iter_150_desvar_bw.png}};
        \node[anchor=south] at (0mm,7mm) {\footnotesize Mathematical design variable $\designvar$};
        \node[anchor=east] at (-33mm,8mm) {\footnotesize (a)};
    \end{tikzpicture}\\[9pt]
    \begin{tikzpicture}[every node/.style={anchor=south west,inner sep=0pt},x=1mm, y=1mm]
        \node[anchor=south] at (0mm,0mm) {
            \includegraphics[scale=0.22,trim={6cm 22.4cm 6cm 0}, clip]
                            {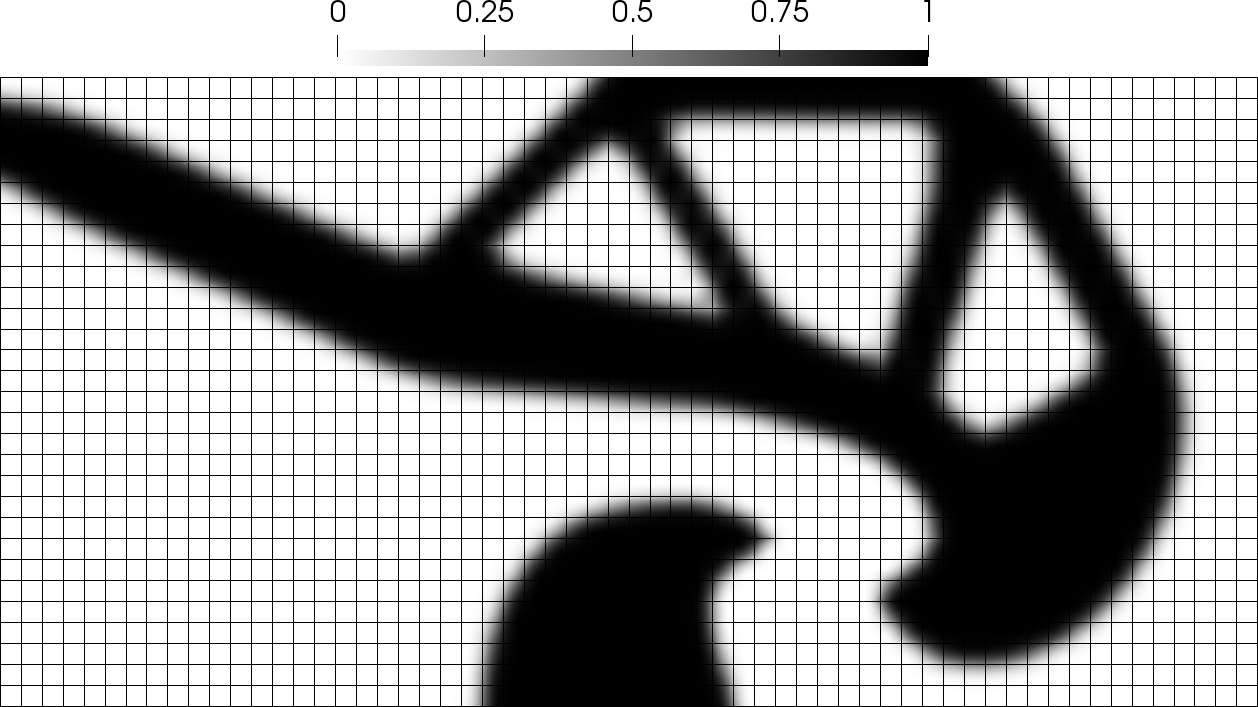}};
        \node[anchor=north] at (0mm,0mm) {
            \includegraphics[scale=0.17,trim={0 0 0 2.5cm}, clip]
                            {fig/iter_150_dens.png}};
        \node[anchor=south] at (0mm,7mm) {\footnotesize Physical density $\rho$};
        \node[anchor=east] at (-33mm,8mm) {\footnotesize (b)};
    \end{tikzpicture}
\caption{Optimization results after 150 iterations. The mathematical design variable field $\designvar$, shown in (a), is interpolated with the bi-linear finite elements chosen for this field, while the physical density field $\rho$, shown in (b) and actually used in the mechanical simulation, is obtained through Eq.~\eqref{eq:projection} without any further post-processing.}
\label{fig:ex1_opt}
\end{figure}

\begin{figure}
    \centering
    \subfloat[Deformed structure at control point 1, $u_{\scriptD x} = 0.1L$.]
             {\includegraphics[scale=0.17, trim={5 360 600 120}, clip]{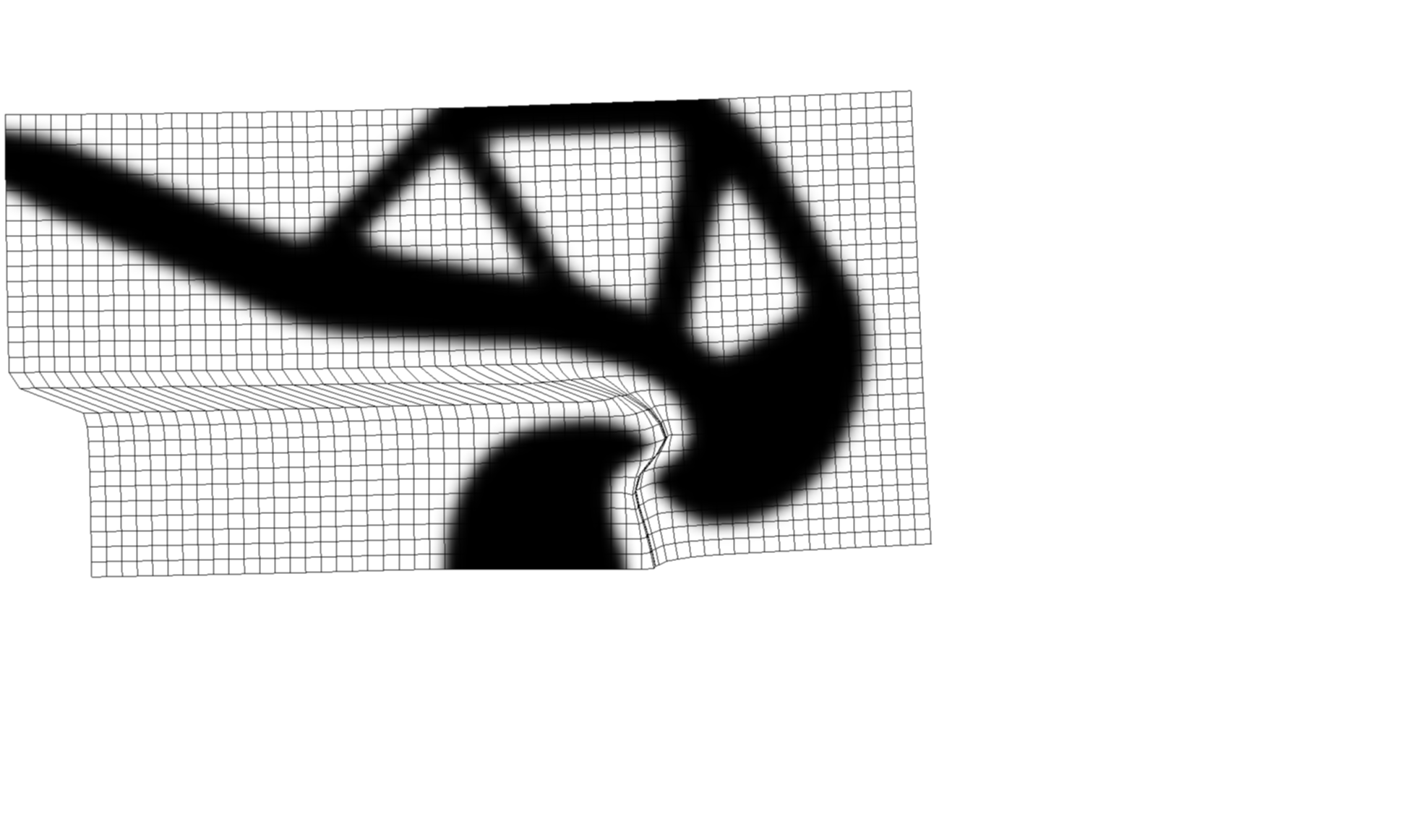}}\\[3pt]
    \subfloat[Deformed structure at control point 2, $u_{\scriptD x} = 0.15L$.]
             {\includegraphics[scale=0.17, trim={5 340 600 80}, clip]{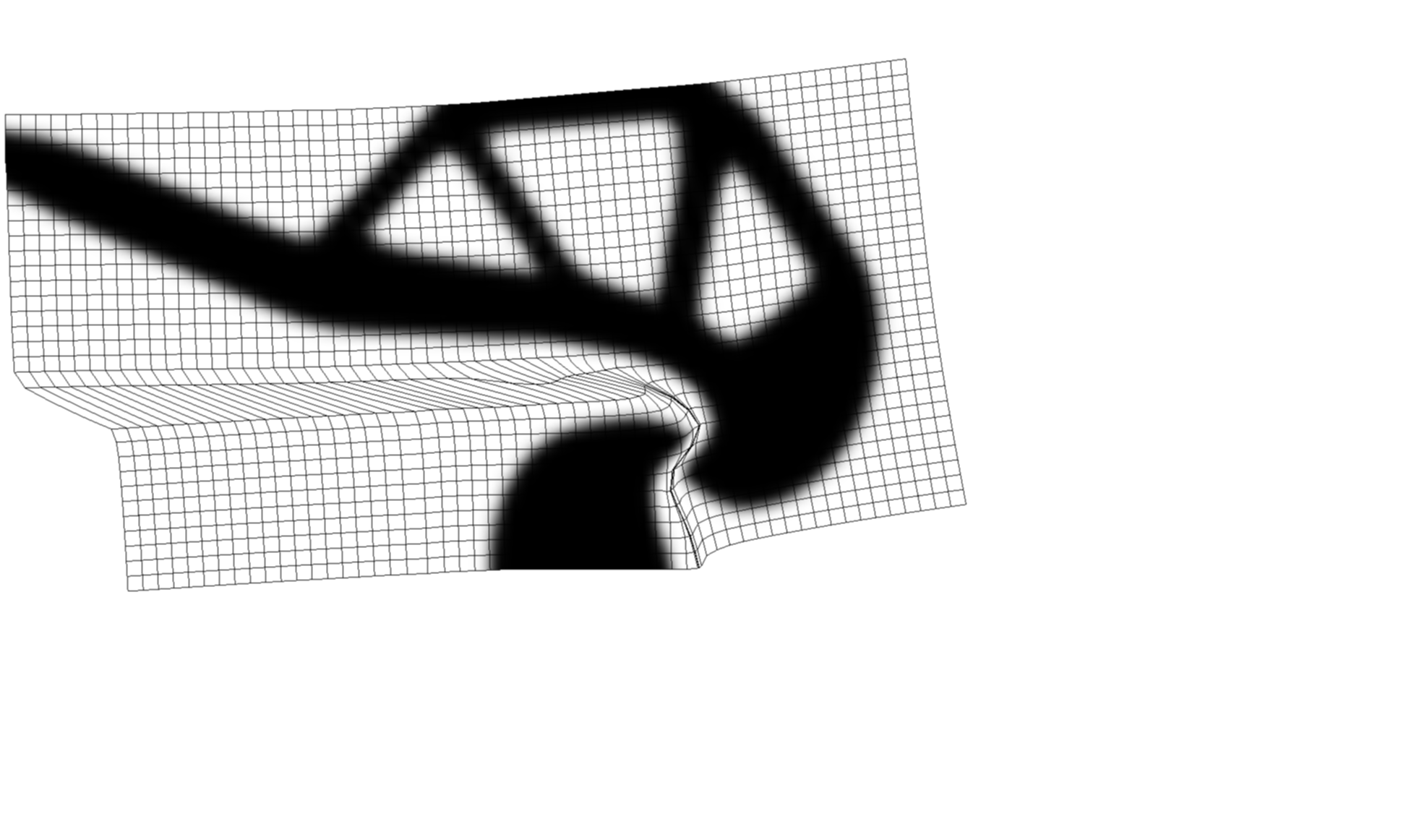}}\\[3pt]
    \subfloat[Deformed post-evaluation structure at $u_{\scriptD x}\!\approx\!0.17L$.
              \label{fig:ex1_posteval_deformed}]
             {\includegraphics[scale=0.17, trim={5 370 600 135}, clip]{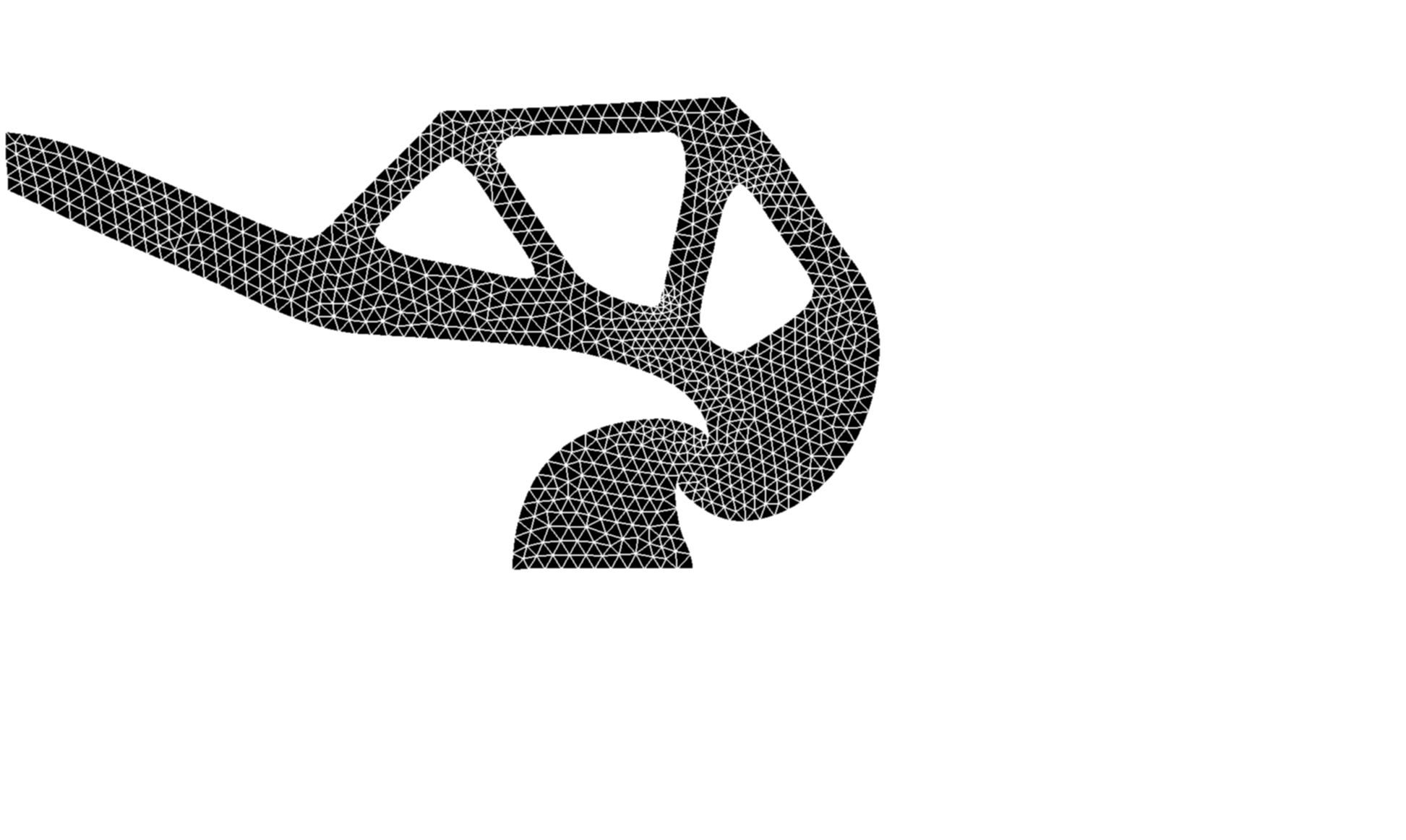}}
\caption{Deformations of the optimized structure at the two control points (a,b). Deformations from post-evaluation with a body fitted mesh (c), based on a discrete solid/void interface at $\rho=0.5$, at the same load as for the optimized structure at control point 2 (marked with an asterisk in Figure~\ref{fig:ex1_load_disp}).}
\label{fig:ex1_u1_u2}
\end{figure}

After 150 design evolution iterations through the objective history shown in Figure~\ref{fig:ex1_iter}, the optimization procedure converges to the design shown in Figure~\ref{fig:ex1_opt}b, in terms of the physical density $\rho$.
The corresponding field of the underlying mathematical design variable $\designvar$ is shown in Figure~\ref{fig:ex1_opt}a.
The performance of the initial and the optimized designs with respect to the set objective can be evaluated based on the corresponding load-displacement curves, shown in Figure~\ref{fig:ex1_load_disp}.
The initial design is very compliant satisfying basically only the zero reaction force required at the first control point but being far from the objective at the second control point.
The optimized design exhibits a small but non-zero reaction force at the first control point and a large reaction force at the second control point, which is nevertheless still below the target traction $\multu^{\scriptnum{2}}_{*x}$.
This solution represents a compromise between the two competing requirements at the two control points.
Unless a design exists that can satisfy both target values, a compromise is to be expected.
The larger, by an order of magnitude, weighting $w^{\scriptnum{1}}$ for the first control point, compared to $w^{\scriptnum{2}}$, results in a smaller deviation from the reaction-free objective at the first point compared to the deviation at the second control point.

As expected, the transition from the approximately decoupled to the stiff response is rather smooth and thus there is no distinct contact point.
This is a consequence of the remaining stiffness in the void between the contact surfaces, and especially in the graded interface, as already has been shown for the C-shape example from Figures~\ref{fig:double_c_contact_cont} and~\ref{fig:cont_rho_load_dist}.
This parasitic interaction, however, is what makes the contact problem differentiable, and therefore more suitable for optimization.

After the first approximately 20 iterations, spent on satisfying the minimum interface width $l_i$ constraint (contained in $\cc_\chi$), which is not respected in the initial guess, the objective function $\cc_\multu$ for the two control points dominates the total objective function $\cc$.
At that point, the objective of the second control point starts to decrease at the cost of an increasing objective for the first control point.
During the whole optimization procedure, the pseudo-time step for each design update increases by several orders of magnitude.
Note that the larger the value of $\Delta t$, the closer Eq.~\eqref{eq:opt_system_damped} is to the exact (local) optimum  from Eq.~\eqref{eq:opt_system}.
The ultimately obtained design contains a truss like structure, typical for compliance minimization, and a hook like coupling, which involves third medium contact.
Figures~\ref{fig:ex1_u1_u2}b and~\ref{fig:ex1_u1_u2}c show the optimized structure in its deformed state at the first and second control point, respectively.
It can be seen both in Figure~\ref{fig:ex1_load_disp} and Figure~\ref{fig:ex1_u1_u2}b, that third medium contact interaction actually occurs prior to the first control point.
Nevertheless, a very clear transition can be observed, in general, from a practically zero initial stiffness to an ultimately stiff post-contact response.

It should be noted that, as the load is increased, the slope of the contact interface in the obtained hook geometry changes due to the deformation of the structure.
Ultimately, at a sufficiently high load, this geometrical change will lead to sliding between the two parts and disengagement.
The large deformation aware optimization procedure, employed in the present work, intrinsically accounts for this effect, compensating for such load-dependent geometrical changes to prevent a disengagement, at least until the second control point.
This is obviously a design characteristic which can only be obtained through a large deformation consistent modeling of both solid and contact mechanics, as applied here.

\subsubsection*{Post-evaluation with body fitted mesh}
The design obtained exhibits a graded solid-void interface of constant width that is perfectly consistent with the prescribed value $l_i$.
Nevertheless, the finite size of the interface is the major source of inaccuracy in the modeling of contact.
For this reason a post-evaluation and comparison with a body fitted mesh is highly relevant.

In order to investigate the behavior of a corresponding discrete void and solid structure, which is actually manufacturable, a body fitted mesh has been generated with GMSH \cite{geuzaine2009a}, based on the interpretation of the contour line $\rho\!=\!0.5$, i.e. $\chi\!=\!0$, as the external surface of the solid.
This body fitted mesh is shown in Figure~\ref{fig:ex1_posteval_deformed}, in its undeformed state, and in Figure~\ref{fig:ex1_posteval_deformed}, in its deformed state under the same load as the graded density structure at the second control point.
Triangular \mbox{6-node} elements and the Lagrange multiplier based contact method from \cite{2015Po-Re} were used for the simulations with this body fitted mesh.
The corresponding load-displacement curve is shown in Figure~\ref{fig:ex1_load_disp} alongside the one from the optimized graded density design.

The post-evaluation curve reveals how delayed the onset of contact actually occurs, being closer to the second control point rather than the first one.
This discrepancy is due to the presence of the graded  density layer between the contacting solids in the graded density structure, seen in Figures~\ref{fig:ex1_u1_u2}a and~\ref{fig:ex1_u1_u2}b.
On the other hand, the post-evaluation confirms a high stiffness of the coupling after onset of contact, which is actually even higher than predicted by the graded density representation, again due to a compliant layer of intermediate density material between the solids in the latter case.
Both effects observed are mainly caused by the remaining stiffness inside the graded solid-void interface, and as such, they are expected to decrease with finer discretization and smaller minimum interface width $l_i$.

All in all, the example presented clearly demonstrates that the proposed method is capable of producing valuable structural designs that exploit internal contact.
These can either be used as input to shape optimization methods or be further optimized within the same framework in a multigrid scheme with subsequent mesh refinements and decreased interface width $l_i$.

\subsubsection*{Additional examples}
\begin{figure}[t!]
    \centering
    \includegraphics[scale=0.17, trim={0 0 0 90},clip]
                    {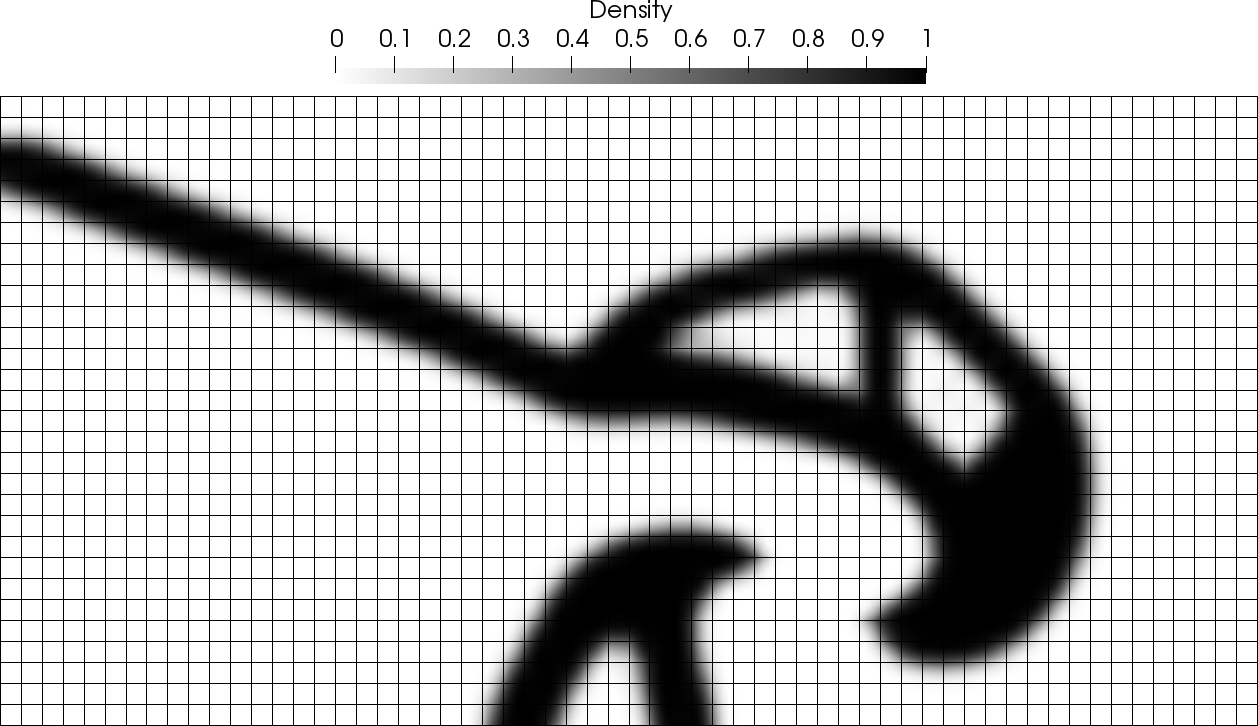}
    \caption{Optimized design (converged $\rho$ field after 150 iterations) for a reduced volume fraction $\bar{\rho}_{\max}\!=\!0.2$ and an accordingly reduced target reaction $q_*^{\scriptnum{2}}\!=\!0.05 \text{MPa}$ for the second control point.}
    \label{fig:ex2_opt}
\end{figure}

\begin{figure}
    \centering
    \includegraphics[scale=0.17, trim={0 0 0 00},clip]
                    {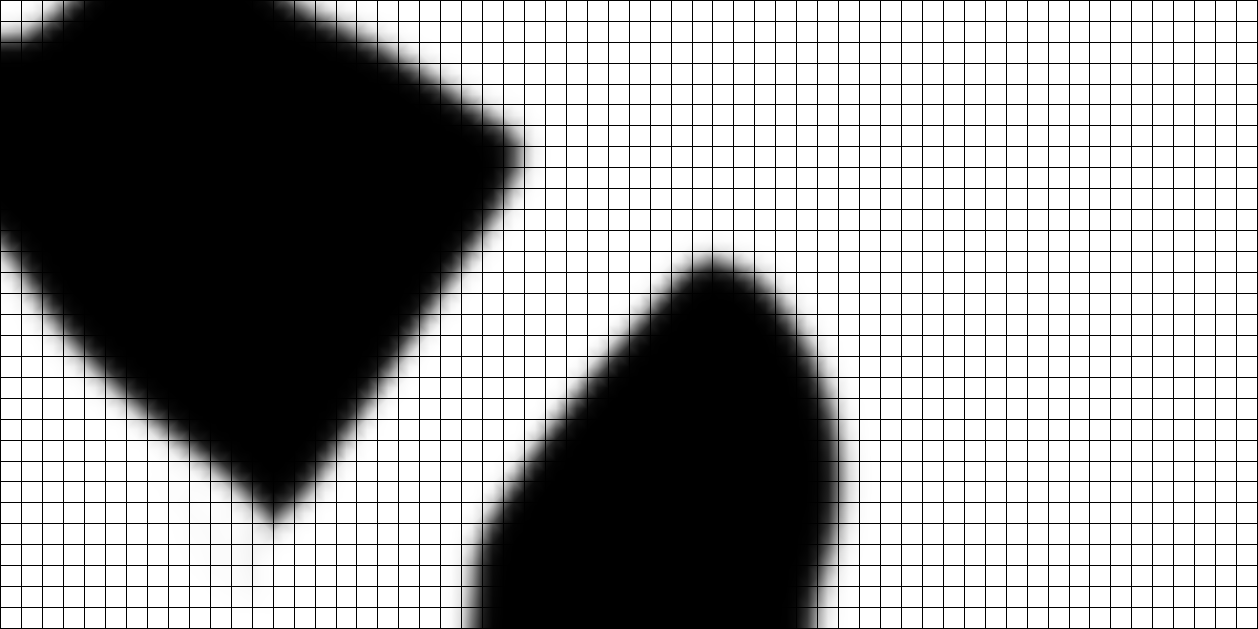}
    \caption{Optimized design (converged $\rho$ field after 300 iterations) for reversed loading direction, i.e. $u_{\scriptD x}^{\scriptnum{1}}\!=\!-0.1 L$, $u_{\scriptD x}^{\scriptnum{2}}\!=\!-0.2 L$ and $q_*^{\scriptnum{2}}\!=\!-0.1 \text{MPa}$, and the original volume fraction $\bar{\rho}_{\max}\!=\!0.35$.}
    \label{fig:ex3_opt}
\end{figure}

To illustrate the generality of the proposed method, two variations of the previous example are presented in Figures~\ref{fig:ex2_opt} and~\ref{fig:ex3_opt}.
For the first variation, the same properties as in Table~\ref{tab:example_params} are used, except for a stricter upper limit on the material usage of $\bar{\rho}_{\max}\!=\!0.2$ and a lower target traction value at the second control point $q_*^{\scriptnum{2}}\!=\!0.05 \text{MPa}$, to account for the reduced volume fraction.
The optimization converged towards the design shown in Figure~\ref{fig:ex2_opt} with similar but thinner features as in the original example.

For the second variation, the loading is reversed by prescribing negative displacements $u_{\scriptD x}^{\scriptnum{1}}\!=\!-0.1 L$ and $u_{\scriptD x}^{\scriptnum{2}}\!=\!-0.2 L$, and desired traction $q_*^{\scriptnum{2}}\!=\!-0.1 \text{MPa}$, while maintaining all further parameters from Table~\ref{tab:example_params}.
Figure~\ref{fig:ex3_opt} shows the optimized design for this variation, after 150 iterations, consisting of two solid blocks, separated by a straight contact interface with orientation perpendicular to the load path between $\Gamma_{\scriptD h}$ and $\Gamma_{\scriptD}$.

\subsubsection*{Computational performance}
For the strongly nonlinear optimization examples of the present work, the computational cost of a single design iteration does not only depend on the number of degrees of freedom of the system, but also on the number of required Newton iterations.
The occurrence of contact does not affect the computational cost for a single Newton iteration.
Nevertheless, the increased nonlinearity due to contact is likely to lead to more frequent time step adaptations according to the scheme of Figure~\ref{fig:sol_alg}.
This will affect the average number of Newton iterations performed per design update.
In quantitative terms, the optimization shown in Figure~\ref{fig:ex1_iter} for the first example of this section, with 46294 degrees of freedom in total, required 26 seconds of cpu time per design update on average, on a single Xeon E5-2660 v3 processor.
The implementation of the example was done in Python with all computational intensive operations, like vector and matrix assemblies performed in the underlying C++ library GetFEM \cite{2021Re-Po}, and linear system solutions performed in the Fortran direct solver MUMPS, \cite{mumps}.

\FloatBarrier
\section{Conclusion}
A method for indirect internal contact in finite strain topology optimization has been presented, based on the third medium contact approach.
Void regions, which are inherently present in density based topology optimization, were exploited as the contact medium, circumventing the need for defining contact surfaces and detecting contact.
To this end, the present work has introduced a new void material regularization which allowed to approximate frictionless contact between structures with discrete or graded solid-void interface, accommodating considerable amounts of sliding.

In general, the usefulness of including a simple and stable contact formulation in topology optimization is twofold.
It enables the optimization procedure to utilize contact as a design feature, as the example shown in the present work, but it also allows to account for collisions, in cases where contact should be avoided.
Hereby it opens up for numerous potential applications in the design of compliant mechanisms, microstructures of periodic material with extremal mechanical properties and energy absorbing structures.

The present work has also assessed the limitations of the proposed approach.
The main source of discrepancies between the proposed third medium contact and graded density design method, and a body fitted mesh based post evaluation, was found to be related to the finite width of the solid-void interface layer.
Possible mitigations have also been discussed.

Besides the contact modeling contribution, this work has also introduced a new general framework for finite strain topology optimization.
The proposed framework is expressed entirely in weak form in a discretization agnostic manner.
It is based on the idea of a simultaneous solving of the fully coupled nonlinear equations for mechanical equilibrium, adjoint analysis, and design optimality, providing a second order consistent optimization scheme.
Moreover, it includes a new method for imposing a length scale for the graded solid-void interface, which is very essential for third medium contact modeling, as explained above.
The potential of this framework has been demonstrated by the optimization examples included, and more in depth studies shall follow in future work.

\begin{acknowledgements}
This work was supported by the Villum Fonden through the Villum investigator project InnoTop.
\end{acknowledgements}

\bibliographystyle{spbasic_unsort}

\bibliography{refs}

\begin{thebibliography}{28}
\providecommand{\natexlab}[1]{#1}
\providecommand{\url}[1]{{#1}}
\providecommand{\urlprefix}{URL }
\expandafter\ifx\csname urlstyle\endcsname\relax
  \providecommand{\doi}[1]{DOI~\discretionary{}{}{}#1}\else
  \providecommand{\doi}{DOI~\discretionary{}{}{}\begingroup
  \urlstyle{rm}\Url}\fi
\providecommand{\eprint}[2][]{\url{#2}}

\bibitem[{Kumar et~al.(2019)Kumar, Saxena, and Sauer}]{kumar2019a}
Kumar P, Saxena A, Sauer RA (2019) Computational synthesis of large deformation
  compliant mechanisms undergoing self and mutual contact. Journal of
  Mechanical Design, Transactions of the Asme 141(1):012302,
  \doi{10.1115/1.4041054}

\bibitem[{Wriggers et~al.(2013)Wriggers, Schröder, and
  Schwarz}]{wriggers2013a}
Wriggers P, Schröder J, Schwarz A (2013) A finite element method for contact
  using a third medium. Computational Mechanics 52(4):837--847,
  \doi{10.1007/s00466-013-0848-5}

\bibitem[{Bog et~al.(2015)Bog, Zander, Kollmannsberger, and
  Rank}]{2015Bo-Za-Ko-Ra}
Bog T, Zander N, Kollmannsberger S, Rank E (2015) Normal contact with high
  order finite elements and a fictitious contact material. Computers \&
  Mathematics with Applications 70(7):1370 -- 1390,
  \doi{10.1016/j.camwa.2015.04.020}

\bibitem[{Oliver et~al.(2009)Oliver, Hartmann, Cante, Weyler, and
  Hernández}]{2009Ol-Ha-Ca-We-He}
Oliver J, Hartmann S, Cante J, Weyler R, Hernández J (2009) A contact domain
  method for large deformation frictional contact problems. part 1: Theoretical
  basis. Computer Methods in Applied Mechanics and Engineering 198(33):2591 --
  2606, \doi{10.1016/j.cma.2009.03.006}

\bibitem[{Simo and Laursen(1992)}]{1992Si-La}
Simo J, Laursen T (1992) An augmented lagrangian treatment of contact problems
  involving friction. Computers \& Structures 42(1):97 -- 116,
  \doi{10.1016/0045-7949(92)90540-G}

\bibitem[{Gitterle et~al.(2010)Gitterle, Popp, Gee, and Wall}]{2010Gi-Po-Ge-Wa}
Gitterle M, Popp A, Gee MW, Wall WA (2010) Finite deformation frictional mortar
  contact using a semi-smooth newton method with consistent linearization.
  International Journal for Numerical Methods in Engineering 84(5):543--571,
  \doi{10.1002/nme.2907}

\bibitem[{Hilding et~al.(1999)Hilding, Klarbring, and Petersson}]{1999Hi-Kl-Pe}
Hilding D, Klarbring A, Petersson J (1999) {Optimization of Structures in
  Unilateral Contact}. Applied Mechanics Reviews 52(4):139--160,
  \doi{10.1115/1.3098931}

\bibitem[{Strömberg(2010)}]{2010St}
Strömberg N (2010) Topology optimization of structures with manufacturing and
  unilateral contact constraints by minimizing an adjustable compliance-volume
  product. Structural and Multidisciplinary Optimization 42(3):341--350,
  \doi{10.1007/s00158-010-0502-1}

\bibitem[{Str{\"o}mberg and Klarbring(2010)}]{2010St-Kl}
Str{\"o}mberg N, Klarbring A (2010) Topology optimization of structures in
  unilateral contact. Structural and Multidisciplinary Optimization
  41(1):57--64, \doi{10.1007/s00158-009-0407-z}

\bibitem[{Str{\"o}mberg(2013)}]{2013St}
Str{\"o}mberg N (2013) The Influence of Sliding Friction on Optimal Topologies,
  Springer Berlin Heidelberg, Berlin, Heidelberg, pp 327--336.
  \doi{10.1007/978-3-642-33968-4_20}

\bibitem[{Kristiansen et~al.(2020)Kristiansen, Poulios, and
  Aage}]{2020Kr-Po-Aa}
Kristiansen H, Poulios K, Aage N (2020) Topology optimization for compliance
  and contact pressure distribution in structural problems with friction.
  Computer Methods in Applied Mechanics and Engineering 364:112915,
  \doi{10.1016/j.cma.2020.112915}

\bibitem[{Luo et~al.(2016)Luo, Li, and Kang}]{2016Lu-Li-Ka}
Luo Y, Li M, Kang Z (2016) Topology optimization of hyperelastic structures
  with frictionless contact supports. International Journal of Solids and
  Structures 81:373 -- 382, \doi{10.1016/j.ijsolstr.2015.12.018}

\bibitem[{Fernandez et~al.(2020)Fernandez, Puso, Solberg, and
  Tortorelli}]{2020Fe-Pu-So-To}
Fernandez F, Puso MA, Solberg J, Tortorelli DA (2020) Topology optimization of
  multiple deformable bodies in contact with large deformations. Computer
  Methods in Applied Mechanics and Engineering 371:113288,
  \doi{10.1016/j.cma.2020.113288}

\bibitem[{Desmorat(2007)}]{2007De}
Desmorat B (2007) Structural rigidity optimization with frictionless unilateral
  contact. International Journal of Solids and Structures 44(3):1132 -- 1144,
  \doi{10.1016/j.ijsolstr.2006.06.010}

\bibitem[{Niu et~al.(2020)Niu, Zhang, and Gao}]{2020Ni-Zh-Ga}
Niu C, Zhang W, Gao T (2020) Topology optimization of elastic contact problems
  with friction using efficient adjoint sensitivity analysis with load
  increment reduction. Computers \& Structures 238:106296,
  \doi{10.1016/j.compstruc.2020.106296}

\bibitem[{Lawry and Maute(2015)}]{2015La-Ma}
Lawry M, Maute K (2015) Level set topology optimization of problems with
  sliding contact interfaces. Structural and Multidisciplinary Optimization
  52(6):1107--1119, \doi{10.1007/s00158-015-1301-5}

\bibitem[{Yoon and Kim(2005)}]{yoon2005a}
Yoon GH, Kim YY (2005) Element connectivity parameterization for topology
  optimization of geometrically nonlinear structures. International Journal of
  Solids and Structures 42(7):1983--2009, \doi{10.1016/j.ijsolstr.2004.09.005}

\bibitem[{Wang et~al.(2014)Wang, Lazarov, Sigmund, and
  Jensen}]{2014Wa-La-Si-Je}
Wang F, Lazarov BS, Sigmund O, Jensen JS (2014) Interpolation scheme for
  fictitious domain techniques and topology optimization of finite strain
  elastic problems. Computer Methods in Applied Mechanics and Engineering
  276:453--472, \doi{10.1016/j.cma.2014.03.021}

\bibitem[{Wallin and Ristinmaa(2015)}]{2015Wa-Ri}
Wallin M, Ristinmaa M (2015) Finite strain topology optimization based on
  phase-field regularization. Structural and Multidisciplinary Optimization
  51(2):305--317, \doi{10.1007/s00158-014-1141-8}

\bibitem[{Cho and Jung(2003)}]{cho2003a}
Cho S, Jung HS (2003) Design sensitivity analysis and topology optimization of
  displacement-loaded non-linear structures. Computer Methods in Applied
  Mechanics and Engineering 192(22-23):2539--2553,
  \doi{10.1016/S0045-7825(03)00274-3}

\bibitem[{Simo et~al.(1985)Simo, Taylor, and Pister}]{simo1985a}
Simo JC, Taylor RL, Pister KS (1985) Variational and projection methods for the
  volume constraint in finite deformation elasto-plasticity. Computer Methods
  in Applied Mechanics and Engineering 51(1-3):177--208, 177--208,
  \doi{10.1016/0045-7825(85)90033-7}

\bibitem[{Poulios and Renard(2015)}]{2015Po-Re}
Poulios K, Renard Y (2015) An unconstrained integral approximation of large
  sliding frictional contact between deformable solids. Computers \& Structures
  153:75 -- 90, \doi{10.1016/j.compstruc.2015.02.027}

\bibitem[{Stolpe and Svanberg(2001)}]{stolpe2001a}
Stolpe M, Svanberg K (2001) An alternative interpolation scheme for minimum
  compliance topology optimization. Structural and Multidisciplinary
  Optimization 22(2):116--124, \doi{10.1007/s001580100129}

\bibitem[{Ebert et~al.(2003)Ebert, Musgrave, Peachey, Perlin, Worley, Mark, and
  Hart}]{2003Eb-Mu-Pe-Pe-Wo-Ma-Ha}
Ebert DS, Musgrave FK, Peachey D, Perlin K, Worley S, Mark WR, Hart JC (2003)
  Texturing and Modeling: A Procedural Approach: Third Edition. Elsevier Inc.

\bibitem[{Klarbring and Torstenfelt(2009)}]{2009Kl-To}
Klarbring A, Torstenfelt B (2009) Ode approach to topology optimization

\bibitem[{Renard and Poulios(2020)}]{2021Re-Po}
Renard Y, Poulios K (2020) Getfem: Automated fe modeling of multiphysics
  problems based on a generic weak form language. ACM Trans Math Softw 47(1),
  \doi{10.1145/3412849}

\bibitem[{Geuzaine and Remacle(2009)}]{geuzaine2009a}
Geuzaine C, Remacle JF (2009) Gmsh: A {3-D} finite element mesh generator with
  built-in pre- and post-processing facilities. International Journal for
  Numerical Methods in Engineering 79(11):1309--1331, \doi{10.1002/nme.2579}

\bibitem[{Amestoy et~al.(2000)Amestoy, Duff, and L'Excellent}]{mumps}
Amestoy P, Duff I, L'Excellent JY (2000) Multifrontal parallel distributed
  symmetric and unsymmetric solvers. Computer Methods in Applied Mechanics and
  Engineering 184(2):501--520, \doi{10.1016/S0045-7825(99)00242-X}

\end{thebibliography}

\end{document}